\documentclass[english,aps,prb,reprint,floatfix,raggedbottom]{revtex4-1}
\usepackage[T1]{fontenc}
\usepackage[latin9]{inputenc}
\usepackage{amsmath}
\usepackage{amssymb}
\usepackage{graphicx}
\usepackage{esint}

\makeatletter

\usepackage[citecolor=blue,linkcolor=blue,colorlinks=true]{hyperref}

\usepackage{subfig}
\usepackage{multirow}
\usepackage{rotating}
\usepackage{array}
\usepackage[table,usenames,dvipsnames]{xcolor}
\makeatother

\definecolor{light-gray}{gray}{0.95}

\usepackage{babel}

\begin{document}

\title{Effects of the interfacial polarization on tunneling in surface coupled quantum dots}

\author{Kuljit S. Virk}

\email{kv2212@columbia.edu}

\affiliation{Department of Chemistry, Columbia University, 3000 Broadway, New York 10027, USA}

\author{David R. Reichman}

\email{drr2103@columubia.edu}

\affiliation{Department of Chemistry, Columbia University, 3000 Broadway, New York 10027, USA}

\author{Mark S. Hybertsen}

\email{mhyberts@bnl.gov}

\affiliation{Center for Functional Nanomaterials, Brookhaven National 
Laboratory, Upton, New York 11973, USA}
\begin{abstract}
Polarization effects are included exactly in a model for a quantum dot
in close proximity to a planar interface.
Efficient incorporation of this potential into the Schr\"{o}dinger equation
is utilized to map out the influence of the image potential effects 
on carrier tunneling in such heterostructures. In particular, the interplay between carrier mass and the dielectric constants of a quantum dot, its surrounding matrix, and the electrode is studied. We find that the polarizability of the planar electrode structure
can significantly increase the tunneling rates for heavier carriers, potentially resulting in a qualitative change in the dependence of tunneling rate on mass.  Our method for treating polarization can be generalized to the screening of two particle interactions, and can thus be applied to calculations such as exciton dissociation and the Coulomb blockade. In contrast to tunneling via intermediate surface localized states of the quantum dot, our work identifies the parameter space over which volume states undergo significant modification in their tunneling characteristics. 
\end{abstract}
\maketitle

\section{Introduction}

The interface between a semiconductor quantum dot (QD) and an extended film is an important part of the functionality of many nanostructured devices, with such diverse applications as photovoltaics~\cite{Nozik2002115,doi:10.1021/jp806791s,xyz}, low threshold lasers~\cite{mazur2,chuang}, single electron memory~\cite{0268-1242-26-1-014026}, and single photon emitters~\cite{Shields:2007ve,0034-4885-68-5-R04}. Understanding the physics of this interface is important for the ultimate technological challenges in these applications: increasing energy conversion efficiency of solar cells, controlling speed and volatility in memory devices, and creating reliable on-demand sources of single photons. 

In novel photovoltaic concepts, absorption of light by QD chromophores can be followed by the transfer of hot carriers to a nearby electrode. This allows the excess carrier energy to be harvested and can fundamentally change solar cell efficiency. Hot carrier transfer was recently demonstrated for optically excited PbSe QDs coupled to  a $\mbox{TiO}_2$ substrate~\cite{xyz}. In laser applications, the reverse process is utilized to design the transfer of cold carriers from a quantum well to a QD, reducing the energy deposition by hot carriers and leading to  higher gain and lower thresholds~\cite{mazur2,chuang}. In addition, rapid progress is  being made on creating novel memory devices based on QDs. The large tunability of QDs allows flexibility to construct fast volatile memories as well as non-volatile memories with very long retention times, low electrical damage, and the possibility of charge storage based on holes as opposed to electrons~\cite{0268-1242-26-1-014026}. The design strategy of each of these devices must take into account the effects of the local electrostatic environment on tunneling.

Optically driven QD emitters are designed such that carrier tunneling and non-radiative energy transfer mechanisms are slower than the exciton radiative recombination rate in order to utilize their fluorescence for single photon generation~\cite{eisaman:071101}. The reverse process of non-radiative energy transfer from a quantum well to colloidal QDs has been also been demonstrated as a way of electrically injecting electron-hole pairs into the dot, where they recombine and emit light~\cite{Achermann:2004zr}. Other electrically driven emitters exploit the tunneling of carriers into and out of epitaxial QDs at two different energies to drive the system into electroluminescence~\cite{Yuan04012002,eisaman:071101,zwiller:1509}. The local electrostatic environment in both types of emitters plays an important role in determining the energy levels and tunneling rates, which in turn control the frequency of emitted light and the efficiency of these devices. This is also becoming more important as research progresses towards single QD emitters~\cite{Shields:2007ve,0034-4885-68-5-R04}.


The tunnel coupling that controls the rate of carrier transfer
in all these tunneling based devices  can be heavily influenced by the image charge effects in the barrier separating the two components between which the carrier transfer takes place. That this classical force on a non-classical electron in the barrier region is essential for accurately determining tunneling rates is well-known in the field of scanning tunneling microscopy~\cite{Binnig1984}. Even when tunneling is not an issue, as in optically driven QD emitters, image charge effects are still important, and serve to lower the exciton energy in the QD, as well as to modify its transition dipole moment. This affects non-radiative energy transfer via the transition amplitudes as well as via resonance with surface excitations that absorb or emit this energy. Therefore, electrostatics within and at the interfaces of sub-systems can be one of the key players in each of the above applications of nanostructures.

In this paper, we re-examine a model system consisting of a spherical QD
coupled to a planar surface.  An exact solution of the Poisson equation,
including all image effects induced by the carrier charge distribution, is presented
in a form that can be efficiently incorporated into
the solution of the Schr\"{o}dinger equation.  
Because of this efficiency, a large parameter space can be explored in this model
and we exploit this capability to demonstrate the impact of the image effects on the carrier tunneling rates.
In addition to barrier lowering effects, we find interesting consequences of the 
\emph{interplay} between the induced electrostatic potential and the confinement of the particle inside the QD. 

\begin{figure}[htb]
\includegraphics[height=1.7in]{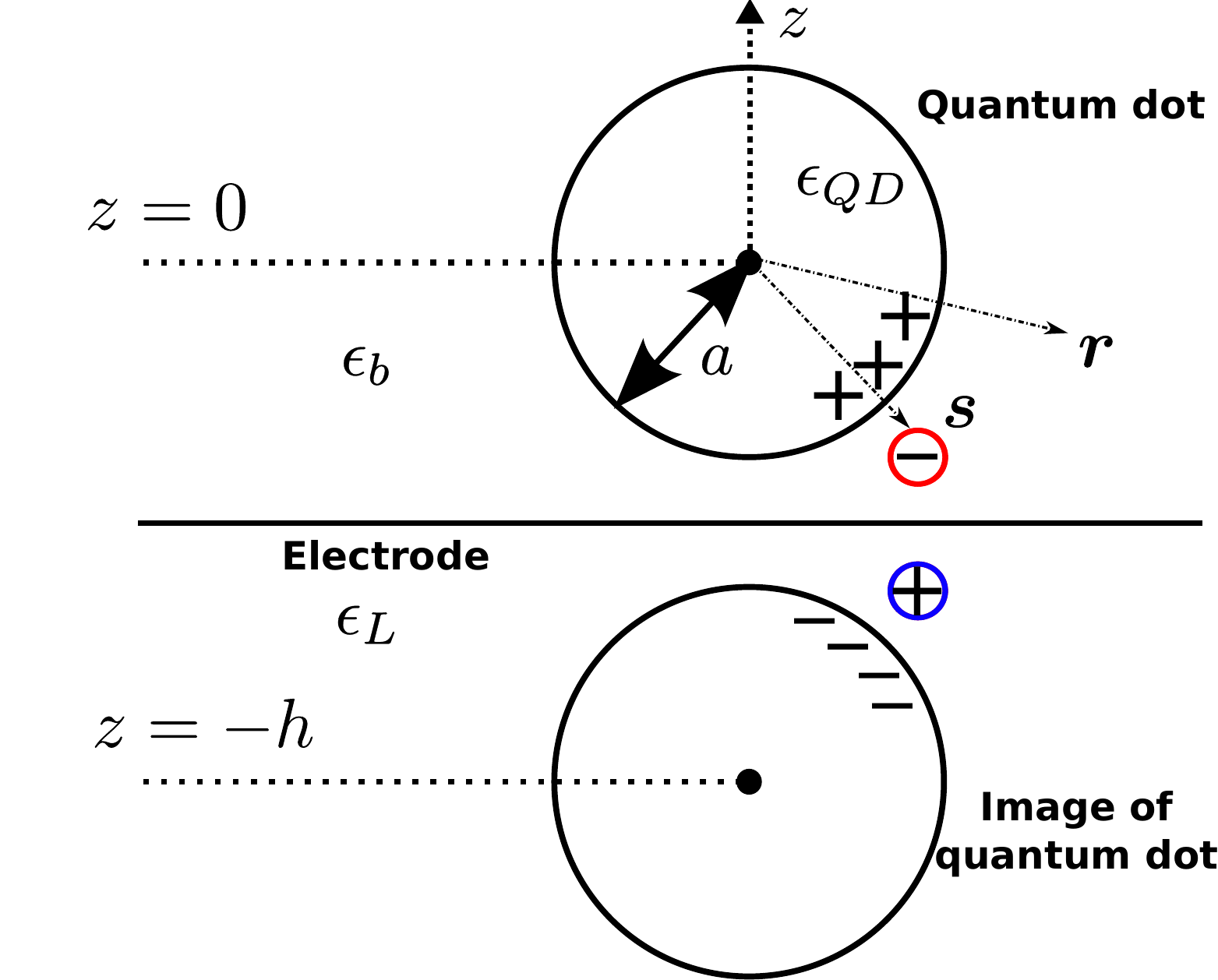}
\vskip 1cm
\includegraphics[height=1.7in]{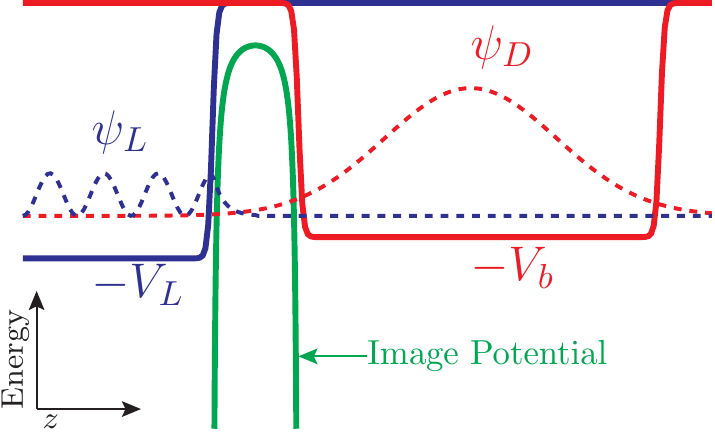}
\caption{\label{fig:drawing} (Left) The physical setup showing coordinates, dielectric constants, and induced charges due to a point charge (red) at position $\boldsymbol{s}$, and its image (blue). (Right) The confinement potentials and schematically drawn image potential and wavefunctions in the QD and the electrode region. Potential is calculated at a general point $\boldsymbol{r}$, and we let $\boldsymbol{r}\rightarrow\boldsymbol{s}$ to obtain the self energy.}
\end{figure}

The system we have modeled is sketched in Fig.~\ref{fig:drawing}, and is represented by a point in the parameter space defined by three dielectric constants (see figure),  confinement potentials, and effective masses, in  each region. To these material parameters, we add QD radius and its distance from the electrode as two geometrical parameters . We find that this space is best understood when we replace $\epsilon_{QD}$  by the ratio $\epsilon=\epsilon_{QD}/\epsilon_b$, and $\epsilon_L$ by the  contrast $f_I = (\epsilon_b-\epsilon_L)/(\epsilon_b+\epsilon_L)$. Thus $f_I=0$ represents no image potential at the electrode interface, and $f_I=-1$ represents the opposite limit of the image potential due to a metal. Similarly $\epsilon=1$ represents the lack of image potential at the QD surface, while $\epsilon\rightarrow\infty$ would represent a metallic nanoparticle as far as electrostatics is concerned.

\begin{figure}[htb]
\includegraphics[width=4in]{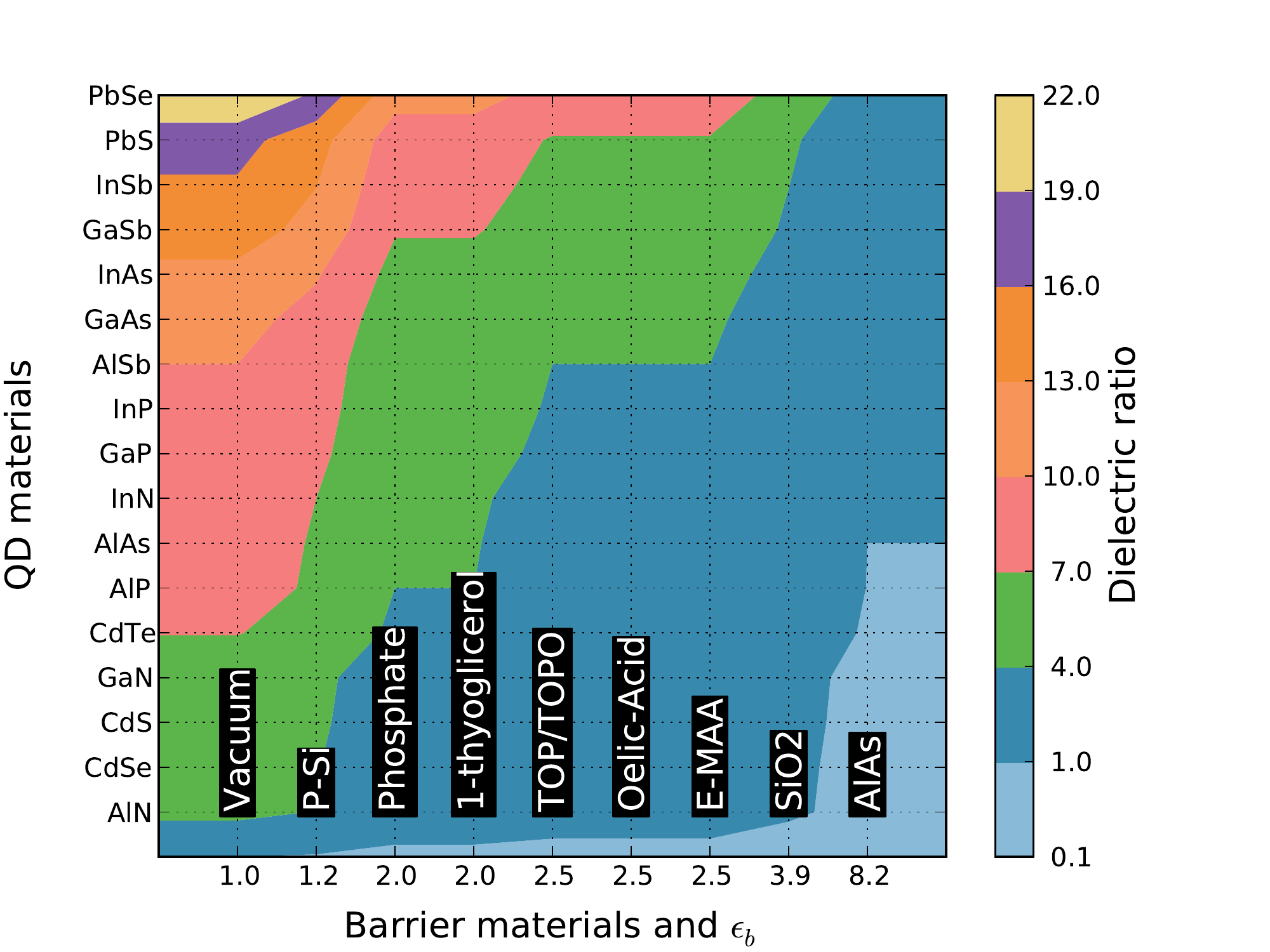}
\caption{\label{fig:materials}The barrier dielectric, $\epsilon_b$, and the ratio, $\epsilon=\epsilon_{QD}/\epsilon_b$, over the space spanned by QD materials on the vertical and barrier materials on the horizontal axis. The values of the barrier dielectric are indicated below the respective materials. At the intersection of two lines, the color indicates the value $\epsilon$ takes for the given combination of QD material and the barrier material the dot is embedded in. The values of dielectric constants used are for $\omega\rightarrow\infty$ since tunneling times are assumed instantaneous.}
\end{figure}

Having defined the strategy of our study in terms of a parameter space, we now connect the most important parameters to real materials and device systems. In Fig.~\ref{fig:materials}, we show a map of the values of $\epsilon_b$ and $\epsilon$ over a space spanned by the QD and the barrier materials. Thus for a given choice of these two parameters in the calculations below, a possible set of materials realizing it can be read from the figure. One first identifies the vertical line closest to $\epsilon_b$ to choose a barrier material. The range of QD materials then fall within the overlap of this line with the color representing $\epsilon$.

In Table~\ref{systemtable}, we categorize the aforementioned classes of devices in terms of the ratio $\epsilon$. Within each class, devices can be found over the entire range of values of the contrast, $f_I$. We present calculations at the two extremes, $f_I=0$ and $f_I=-1$, for each system studied below. However the choice of $f_I$ for a given electrode can be subtle. In systems with low $\epsilon_b$, $f_I\approx -1$ for a substrate made of metal or an undoped semiconductor. At high $\epsilon_b$, as in the case of III-V epitaxial systems, $f_I\approx 0$ for an undoped substrate while it may be $-1$ for metallic electrodes. In the latter case, it is important to realize that the metallic electrodes are often doped semiconductors in which the carrier densities are of the order of $10^{18}$ cm$^{-3}$, which yields an average distance between dopants equal to 100 \AA. For QDs with diameters smaller than 10 nm, the electrode may be best approximated via the static dielectric constant of its host material. Thus setting the value of the parameter $f_I$  can be subtle in some systems, and its choice must take into account the properties such as dopant densities and the plasma frequency inside the electrode material.

\begin{table}[htb]
{\small

\begin{center}

\begin{tabular}{p{1\columnwidth}}

\cline{1-1}\noalign{\smallskip}

{\cellcolor{light-gray}{\bf low $\epsilon$:} III-V and II-VI systems of epitaxial QD layers} \\

     \\ 
 optically and electrically pumped emitters \\
 volatile single electron memory \\
 QD lasers\\
     \\

\hline\noalign{\smallskip}

 {\cellcolor{light-gray}
\parbox{1\columnwidth}{\raggedright{\bf high $\epsilon$:} III-V, II-VI, Group IV colloidal QDs in vacuum, 
organic, p-Si, $\mbox{SiO}_2$, and MgS matrixes}}\\
\\
 optically and electrically pumped emitters \\
 energy transfer pumped emitters \\
 non-volatile single electron memory\\
  nanocrystal based photovoltaics\\ 
     \\

\hline

\end{tabular}
\caption{\label{systemtable}Various QD-based systems in the parameter space defined by the dielectric ratio, $\epsilon$, between the barrier and the QD. The classification, and the shaded heading in the low and high $\epsilon$ blocks refers to the most common types of systems. The nanocrystal based systems may be found with both high and low $\epsilon$ depending on the materials used. Specific devices realized under each system often span the entire range of dielectric contrast $f_I$ between the background and the planar medium.}
\end{center}
}

\end{table}

Let us consider the device examples of Table~\ref{systemtable} in light of the parameter space of our model. The physical quantities important in the interplay of confinement and the induced potential are the effective mass inside the QD, the barrier potential, and the barrier dielectric. This implies a dependence of tunneling rate on the mass \emph{inside} the QD, which may at first seem strange  since a Fermi Golden Rule calculation of the tunneling rate involves only the wavefunctions within the barrier region~\cite{bardeen}; the dependence of this rate on the mass inside the \emph{barrier} is well-known \cite{chuang}. However, the amplitude of the wavefunction in the barrier region depends on the behavior of the wavefunction inside the QD and at its boundaries, and therefore
 on the mass inside the QD volume. The substantial difference in effective mass often seen in semiconductors (including electrons, heavy holes and light holes) then suggests that the rates for electron and hole tunneling in QD-based structures could also be quite different, as seen for example in models for QD laser structures \cite{chuang}. 
 
 In fact, depending on the dielectric contrast between the barrier and the QD, we find that image effects can lead to substantial localization of heavy carriers inside the QD with a significant change in tunneling rate. The trend in tunneling rate with carrier mass can thus be substantially altered from the simpler picture found when image effects are neglected. In particular, we identify a regime in which a heavy carrier tunnels at a similar or much higher rate than a light carrier. Interestingly Mazur \emph{et al} point to several experimental measurements in QD laser systems showing more modest differences between these rates for the two carrier types\cite{mazur2} than is expected from a simple WKB~\cite{merzbacher} analysis~\cite{chuang}. While the interpretation of these particular experiments requires a proper accounting of surface states and defect states within the barrier, we find that in a parameter regime somewhat different from the conditions in these experiments, the image effects can also yield similar results.

As is known from elementary electrostatics, the polarization induced at an interface of two materials is directly proportional to the dielectric contrast between them~\cite{jackson-book}. In the experiment of Tisdale \emph{et al.}~\cite{xyz}, the dielectric constant of the $\mbox{TiO}_2$ electrode is 8.4 and that of the PbSe QD is 22.4 at high frequencies~\cite{springer-materials}. On the other hand, the organic materials used to passivate the QDs, hydrazine and ethanedithiol, both have optical refractive index of 1.5~\cite{springer-materials}. Thus approximating the high-frequency dielectric constant ($\epsilon_\infty$) within the barrier region to be 2.3, substantial image effects are expected at both the electrode and the QD surface. In our model, this corresponds to high $\epsilon$ and $f_I\rightarrow -1$. We find that this system should be in the parameter regime in which heavier particles tunnel faster than the lighter ones. 

Furthermore, the most interesting materials for the QD chromophore in photovoltaic systems generally have small bandgaps, and therefore large dielectric constants. These are generally colloidal QDs passivated by organic materials with relatively low dielectric constants at high frequencies. Therefore, these systems lie in a regime where the effects of interfacial polarization are maximized. A range of behaviors can thus result from the interplay of confinement due to the barrier potential and the QD dielectric constant relative to that in the barrier.

Laser systems based only on III-V semiconductors are in a region where the dielectric contrast between each pair of materials is low, and thus the interfacial polarization is small. In our model, this corresponds to low $\epsilon$ and $f_I\rightarrow 0$. On the other hand, non-volatile single electron memory systems generally contain narrow gap nanocrystals (\emph{e.g.} Si, InAs, GaSb) embedded inside wide gap layers (\emph{e.g.} III-nitrides, AlAs, silica) for greater reliability. Therefore they lie in a regime where the interfacial polarization may be large, much like that for the photovoltaic systems discussed above. However, unlike photovoltaic systems, these devices often have tunable potentials, with barriers that are often thicker than 10 nm. Therefore only the image effects at the QD/barrier interface are important, or in terms of our model parameters: a high $\epsilon$ and $f_I\rightarrow 0$.

Quantum dot emitters are often embedded in high dielectric materials, and therefore have little dielectric contrast with their surroundings. This limits the light collection efficiency, and there is thus a growing interest in systems with QDs coupled to optical fibers~\cite{davanco:121101}, or embedded in host materials such as photonic crystals~\cite{vuckovic:2374}, MgS~\cite{PhysRevB.73.245308}, and porous silicon~\cite{richter:142107}. In addition, large confinement potentials in these cutting edge systems would also remove the limitation of operating only at low temperatures. From the foregoing discussion, and the details presented below, these types of systems are  expected to show large interfacial polarization at the QD boundary, which corresponds to  a high $\epsilon$ in our model. Furthermore $f_I\approx 0$ for optically pumped emitters since they do not depend on a nearby charge reservoir. In electrically pumped emitters, $f_I$ may take values between 0 and $-1$, but the electrode properties must first be carefully examined  to set a particular value as we emphasized in the above discussion. In novel energy transfer emitters~\cite{Achermann:2004zr}, passivated colloidal QDs lie on the surface of a dielectric, resulting in a value of $-1<f_I<0$. Thus we expect QD emitters to span a large region in the parameter space of our model, and to exhibit a rich interplay between polarization and confinement affecting the binding energy, fluoresence, and electroluminescence in these systems.

Our paper is organized as follows. In Sec.~\ref{sec:theory}, we begin with the Schr\"{o}dinger equation for a single electron inside the system shown in Fig.~\ref{fig:drawing}(a), and derive explicit expressions for its potential energy, and its rate of tunneling between the QD and the electrode. The mathematical details of the derivation are discussed in the Appendix.  In Sec.~\ref{sec:results}, we apply this theory to computationally explore the parameter space in the manner described above. In particular, we present the crossover of trends in tunneling rates for various combinations of parameters, and interpret this behavior in terms of the  interplay discussed above. Following this section are our conclusions and remarks on the use of our results inside more sophisticated \emph{ab-initio} calculations of coupled Schr\"{o}dinger and Poisson equations.


\section{Theory}\label{sec:theory}
\subsection{Electronic states}
The basic features of many devices can be modeled with the QD treated as a dielectric sphere 
and with the electronic structure of the device treated via an effective mass approximation.
For a one-band model \cite{banyai},
\begin{equation}
\left[-\frac{\hbar^{2}}{2m_0}\nabla\cdot\frac{1}{m(\boldsymbol{r})}\nabla
+V(\boldsymbol{r})+\frac{e^{2}\Sigma(\boldsymbol{r})}{4\pi
\varepsilon_{0}\epsilon_{b}a}-E\right]\psi(\boldsymbol{r}) =
0,\label{eq:schrod}
\end{equation}
where $a$ is the QD radius, $\epsilon_{b}$ is the dielectric constant
of the barrier region, $m(\boldsymbol{r})$ is the ratio of effective mass to bare mass $m_0$, $V(\boldsymbol{r})$ is the confinement potential,
and $\Sigma(\boldsymbol{r})$ is the electrostatic self-energy in
the dimensionless form. 

The electrostatic self-energy of a charge in an \emph{isolated} spherical QD was first derived
by Brus\cite{brus:4403} for the infinite spherical well model, and
then by Ban\'yai \emph{et al.} for a finite well \cite{banyai}. Here we generalize these results by accounting for the geometric
progression of image charges induced on the electrode and the QD surface.
The potential energy is derived based on re-expansion of spherical harmonics~\cite{caola} around a shifted origin. 
This approach may be found in several sources, including recent work on plasmonics of nanoparticles supported on a substrate\cite{PhysRevB.83.075426}.  As appropriate for the optical domain, the solutions presented there apply to an electric field uniform over the particle. Solutions to the Laplace equation within the space bounded by axially symmetric charge distributions in the same geometry have also been given \cite{Matsuyama1995}. We present a solution
applicable to an arbitrary charge density, as needed for the self energy in the Schr\"{o}dinger equation. 
 
We first consider the potential
at $\boldsymbol{r}$ of a source charge at $\boldsymbol{s}$, and write it as a superposition of
four sources: the point charge, the multipole moments of the dielectric
sphere, the image of the point charge in the plane, 
and the image of the multipole moments in the plane (see Fig. \ref{fig:drawing}). Subjecting the contribution from surface charge densities to a multipole expansion, we write the potential as
%
%
%
\begin{widetext}
\begin{eqnarray}
\Phi(\boldsymbol{r};\boldsymbol{s})
&=&	\frac{\mathfrak{q}(\boldsymbol{s})e}{4\pi\varepsilon_{0}\epsilon_b\left|\boldsymbol{r}-\boldsymbol{s}\right|}+\frac{f_I\mathfrak{q}(\boldsymbol{s})e}{4\pi\varepsilon_{0}\epsilon_b\left|\boldsymbol{r}-\boldsymbol{s}_I\right|}
		+\frac{e}{\varepsilon_{0}\epsilon_b a}\sum_{l,p}\frac{Q_{lp}(\boldsymbol{s})}{2l+1}\left[\Theta(a-r)\left(\frac{r}{a}\right)^{l}+\Theta(r-a)\left(\frac{a}{r}\right)^{l+1}\right]Y_{lp}(\hat{\boldsymbol{r}}) \nonumber\\
&&
		+\frac{ef_I}{\varepsilon_{0}\epsilon_b a}\sum_{lp}\left(-1\right)^{l+p}\frac{Q_{lp}(\boldsymbol{s})}{2l+1}\left(\frac{a}{\left|\boldsymbol{r}+\boldsymbol{h}\right|}\right)^{l+1}
Y_{lp}\left(\frac{\boldsymbol{r}+\boldsymbol{h}}{\left|\boldsymbol{r}+\boldsymbol{h}\right|}\right),\label{eq:phi}
\end{eqnarray}
\end{widetext}
where $\boldsymbol{s}_I=\boldsymbol{s}-2\boldsymbol{s}\cdot\hat{z}\hat{z}-\boldsymbol{h}$ is the position vector of the image charge. The dimensionless function $\mathfrak{q}(\boldsymbol{s})$ is fixed  by requiring that the total charge in the real half space equal $q$ in units of the elementary charge $e$. 
This function allows us to isolate the effects of dielectric screening of the charge in a continuous medium when the point $\boldsymbol{s}$ lies inside the sphere, and is given by,
\begin{equation}
\mathfrak{q}(\boldsymbol{s})=q\left[\Theta(s-a)+\frac{1}{\epsilon}\Theta(a-s)\right].
\end{equation}
 In the second and third terms of \eqref{eq:phi}, the functions $Y_{lp}(\hat{r})$ are spherical harmonics written as a function of the direction vector.  The second term corresponds to the surface charge density in real space, while the third term is due to the image of this charge density. In this term, we have already accounted for the change in the sign of the multipoles in the image space, which depends on $l$ and $m$, where the integer 
$l$ is  the total angular momentum and $p$ the azimuthal quantum number.

We find the self consistent multipole moments, $Q_{lp}(\boldsymbol{s})$ by enforcing the electrostatic boundary conditions at the surface of the sphere as shown in Appendix~\ref{multipole-moments}. 
Imposing axial symmetry of the complete system uncouples the different $p$, and
the $Q_{lp}(\boldsymbol{s})$ for each $p$ satisfy a set of linear equations indexed by $l$:
\begin{equation}
\boldsymbol{Q}(\boldsymbol{s}) = \mathfrak{q}(\boldsymbol{s})[\boldsymbol{Q}^{(0)}(\boldsymbol{s})+f_I\boldsymbol{Q}^{(0)}(\boldsymbol{s}_I)]+\boldsymbol{A}(a/h)\boldsymbol{Q}(\boldsymbol{s}).
\end{equation}
 The $Q_{lp}^{(0)}(\boldsymbol{s})$ and $Q_{lp}^{(0)}(\boldsymbol{s}_I)$ are induced just by the source point charge and the image of the point
charge respectively, where $f_I=(\epsilon_{b}-\epsilon_{L})/
(\epsilon_{b}+\epsilon_{L})$ (Fig. \ref{fig:drawing}). 
The $Q_{lp}^{(0)}(\boldsymbol{s})$ are given by,%
\begin{eqnarray*}
Q_{lp}^{(0)}(\boldsymbol{s})&=&\frac{\left(\epsilon-1\right)}{\left(\epsilon+1\right)l+1}Y_{lp}^*(\hat{\boldsymbol{s}})\begin{cases}
\left(l+1\right)\left(s/a\right)^{l}, & s<a\\
-l\left(a/s\right)^{l+1}, & s>a
\end{cases},%
\end{eqnarray*}%
where $\epsilon = \epsilon_{QD}/\epsilon_b$.  The matrix elements $A_{ll'}^{p}$ 
depend on the radius of the sphere and the distance between the center of the sphere 
and its image (Fig. \ref{fig:drawing}); they are given by%
\begin{eqnarray}
A_{ll'}^p&=&\frac{f_I\left(2l+1\right)}{\left(\epsilon+1\right)l+1}\gamma_{l'p}\gamma_{lp} \label{eq:A-matrix-elements} \\
&&\times\frac{l
\left(\epsilon-1\right)\left(-1\right)^{l'+l}(l+l')!}
{\sqrt{\left(2l'+1\right)\left(2l+1\right)}l!l'!}\left(\frac{a}{h}\right)^{l
+l'+1},\nonumber
\end{eqnarray}%
where, 
\begin{equation}
\gamma_{np}=\frac{n!}{\sqrt{\left(n-p\right)!\left(n+p\right)!}}.\label{eq:gamma}
\end{equation} 
The details of this entire derivation are also provided in Appendix~\ref{multipole-moments}.
The above equation for self consistent multipole moments couple the moments for all angular momenta $l$ to each other. In practice, truncation must be applied by choosing $l<l_{max}$, and increasing $l_{max}$ until the final result converges. We found that $l_{max}>15$ is sufficient in all cases we have applied this model to, and generally set $l_{max}=50$.

Equations \eqref{eq:phi}-\eqref{eq:gamma} can be applied immediately to the interaction energy of two charged particles, e.g. for treatment of excitons, by considering one charge placed at $\boldsymbol{s}$ and using $Q_{lp}(\boldsymbol{s})$ to calculate the potential at $\boldsymbol{r}$, the location of the second charge. In the following, we focus only on the single particle self-energy $\Sigma$ and its consequences for tunneling. We write $\Sigma$ as a sum of two contributions, $\Sigma_{0}+
\Sigma_{1}$,
where $\Sigma_{0}$ is the interaction energy \emph{without} the planar
surface ($\boldsymbol{\rho}=\boldsymbol{r}/a$), 
\begin{eqnarray}
&&\Sigma_{0}(\boldsymbol{\rho})=\frac{q^2}{2}\left(1-\epsilon^{-1}\right)\Theta(1-\rho)\\
&+&\frac{q^2}{2}\frac{\epsilon-1}{\epsilon+1}\Theta(1-\rho)\left[\frac{\epsilon^{-1}\rho^{2}}{1-\rho^{2}}-\frac{\rho^{-2}\ln\left(1-\rho^{2}\right)+1}{\left(\epsilon+1\right)}\right.\nonumber\\
&&\left.+\frac{\epsilon}{\left(\epsilon+1\right)}\sum_{l=1}^{\infty}\frac{\rho^{2l}}{\left[\left(\epsilon+1\right)l+1\right]\left(l+1\right)}\right]\nonumber\\
&+&\frac{q^2}{2}\frac{\epsilon-1}{\epsilon+1}\Theta(\rho-1)\left[\frac{1}{1-\rho^{2}}-\frac{\ln\left(1-\rho^{-2}\right)+\rho^{-2}}{\left(\epsilon+1\right)}\right.\nonumber\\
&&+\left.\frac{\epsilon \rho^{-2}}{\left(\epsilon+1\right)}\sum_{l=1}^{\infty}\frac{\rho^{-2l}}{\left[\left(\epsilon+1\right)l+1\right]\left(l+1\right)}\right]\nonumber,
\end{eqnarray}
and $\Sigma_{1}$ is the correction due to the electrode surface,
\begin{eqnarray}
& &\Sigma_{1}(\boldsymbol{\rho})=\\
&&2\pi q\sum_{l=1}^{\infty}\sum_{p=-l}^{l}
\frac{Q_{lp}(\boldsymbol{\rho})-Q_{lp}^{(0)}(\boldsymbol{\rho})}{2l+1}
Y{}_{lp}(\hat{\mathbf{\rho}}) S_l(\rho)\nonumber\\
& &+2\pi f_{I}q\sum_{l=1}^{\infty}\sum_{p=-l}^{l}\frac{\left(-1\right)^{l+p}Q_{lp}(\boldsymbol{\rho})}{2l+1}Y_{lp}
(\hat{\mathbf{\rho}})S_l(|\boldsymbol{\rho}+\boldsymbol{h}/a|)\nonumber
\\
& & +\frac{f_Iq^2}{2}\left[\frac{\Theta(\rho-1)+\epsilon^{-1}\Theta(1-\rho)}{(2\boldsymbol{\rho}\cdot\hat{z}
+h/a)}+\frac{\left(1-\epsilon^{-1}\right)\Theta(1-\rho)}{\left|\boldsymbol{\rho}+
\boldsymbol{h}/a\right|}\right].\nonumber
\end{eqnarray}%
Note that in the above expression, both series  start at $l=1$ because the monopole contribution from the induced charge on the image sphere is explicitly taken into account by the last term.
In the derivation of the formula for $\Sigma_0$, we followed B\'anayi \emph{et al}~\cite{banyai} to separate the divergent terms of the multipole series and sum them to identify closed form expressions that can be regularized (see Appendix~\ref{regularize}).  The form of our expressions is slightly different from theirs due to the zeroth order multipole contribution written  explicitly in our formula, while it is subsumed into the series in theirs. It represents the part of energy stored in the dielectric medium of infinite extent when a charge is embedded inside it. Here this energy is for the QD dielectric relative to the barrier medium.  In both the above formulas, $S_l(\rho)$ equals $\rho^{l}$ for $\rho<1$ and $\rho^{-(l+1)}$ otherwise. 

Since macroscopic electrostatics is being used in this model, the material parameters at the QD boundary are taken to follow an abrupt profile. In the case of effective mass, we  handle the discontinuity numerically using the ghost fluid method\cite{Liu2000}. In the case of electrostatic potential, the self energy diverges at the boundary, which could be regularized by the microscopic profile of the potential. Since the present paper does not include defects of the atomic scale potential, we do not compute this potential and instead continuously join the self energy with the confining potentials of the 
electrode and the QD relative to the background medium\cite{Chulkov1999330}.  We do so by  linearly interpolating $\Sigma$ between its computed values within one lattice spacing  inside and outside the QD interface. At the remaining interface, the planar surface, we set the image charge interaction to 
\begin{equation}
V_{img}=af_Iq^2\frac{1-\exp\{-(z+h/2)/\eta\}}{4(z+h/2)}\label{eq:Vimg},
\end{equation}
 where $\eta$ is chosen to yield a specified potential at the image plane. Results are generally insensitive to $\eta$ for sufficiently deep surface potential because we focus only on wavefunctions that are decaying exponentially in this region.

\subsection{Tunneling rates}

We now consider tunneling of a single electron from the QD to the electrode. We model the electrode as a thick square well of finite depth with respect to the barrier at $z=-h/2$, and an infinite potential wall at the opposite end $z=-h/2-w$, letting $w\rightarrow\infty$ in the final calculation to effect a semi-infinite substrate. Thus letting $\boldsymbol{k}$ be the in-plane wavevector and $k_z$ the transverse wavenumber, we write a complete set of basis functions of the electrode as,
\newcommand{\boldk}{\boldsymbol{k}}
\begin{eqnarray}
\psi_{L}(k_z,\boldk;\boldsymbol{r}) &=& A(k_z,k)e^{i\boldsymbol{k}\cdot\boldsymbol{r}}\left[
e^{-\kappa(z+h/2)}\Theta\left(2z+h\right)\right. \nonumber\\
&+&\left.\sin\left(k_z\left(z+h/2\right)+\delta\right)\Theta\left(h-2z\right)\right],
\end{eqnarray}
where  $A$ normalizes $\psi_L$ to the volume of the electrode (in the limit of semi-infinite electrode below, the volume drops out in the final expressions) and $\tan{\delta}=k_z/\kappa$. In this simple model, the numbers $k_z,\kappa$, and $\boldk$ are all real valued, and a state at given energy $E$ satisfies
\begin{eqnarray}
k_z^2 &=& \frac{2m_L (E+V_L)}{\hbar^2}-k^2,\\
\kappa^2 &=& \frac{m_b}{m_L}k^2-\frac{2m_b E}{\hbar^2}\label{eq:kappasq},
\end{eqnarray}
and is admissible only if $k_z^2>0$ and $\kappa^2>0$. In the expressions above, we have introduced $m_L$ as the effective mass inside the electrode. More realistic models, such as those with surface states, will be necessary to obtain accurate rates. Regardless of the model used, the states would still decay exponentially perpendicular to the surface, and would be described by a (lattice) translational symmetry along the electrode surface~\cite{PhysRevB.31.805}. In addition, the states enter into our tunneling rate below in such a way that only the wavefunction outside of the electrode is required; above, we have included the simplified form in $z<-h/2$ region for completeness. Therefore, at the length scale used in the calculations below, the matrix elements between the QD states and the basis states of the electrode, are captured well by the above model. However, significant quantitative differences would arise due changes in density of states, as well as additional modes due to the surface states.  

Having defined the basis set for the final state of the electron escaping the QD, the tunneling rates are computed using the Fermi Golden rule, 
\begin{equation}
\gamma_n = \frac{1}{\hbar}\int_0^\infty dk_z\int d^2\boldk \left|\left<\psi_{L;k_z,\boldk}\right|V\left|\psi_{D;n}\right>\right|^2\delta(E_{k_z,k}-E_n),
\end{equation}
where $E_{k_z,k}=\hbar^2/2m(k_z^2+k^2)$ is the energy of the electrode basis state. The matrix element is calculated using Bardeen tunneling theory, in which it is implicitly assumed that the wavefunctions of the two subsystems are orthogonal. Enforcing this assumption yields,
\begin{eqnarray}
&&\left<\psi_{L;k_z,\boldk}\right|V\left|\psi_{D;n}\right>\nonumber\\
&=&\int d^2\boldsymbol{r}\psi_{L}^{*}
(k_z,\boldk;\boldsymbol{r})\left[H(\boldsymbol{r})-H_{L}(\boldsymbol{r})\right]
\psi_{D;n}(\boldsymbol{r}).\label{eq:transition}
\end{eqnarray} 
Here $H_L$ is the Hamiltonian for the electrode in isolation of the QD, and the eigenstates of which are the above $\psi_{L}(k_z,\boldk;\boldsymbol{r})$. 

The Bardeen tunneling matrix element is traditionally written as a surface integral. This form can be obtained by the use of the divergence theorem, but the formula~\eqref{eq:transition} is more convenient for our calculations. However, the aspect of Bardeen theory that is essential in this expression is the orthogonality assumption, which yields  a matrix element of the difference between energies. We obtained this form by adding and subtracting $H_L$ to $H$, and then exploiting the relation  $\langle\psi_{L;k_z,\boldk}|H_L|\psi_D\rangle=E_{L;k_z,\boldk}\langle\psi_{L}|\psi_D\rangle\approx0$. The subtraction is advantageous because it naturally restricts the integration domain in \eqref{eq:transition} to the volume outside the electrode, and thus allows the possibility of computing $\psi_D$ and $\psi_L$ in separate calculations, in which one would focus on the QD and electrode properties respectively. The subtracted Hamiltonian $H_L$ is non-vanishing in calculations in which the termination of the potential at the surface occurs smoothly into the junction, as is the case in atomic scale modeling. In the present calculation, $H_L$ describes a square well and thus makes no contribution outside the planar surface. 

From \eqref{eq:kappasq}, we see that $\kappa\propto\sqrt{m_b}$, which in turn is proportional to $\log{\gamma}$. Remarkably, we find that the tunneling rate also shows a similar dependence on the mass \emph{inside} the QD. While this is not a fundamental relationship, we find that when the tunneling rates follow a monotonically decaying profile with respect to $m$, they can be described well by the expression,
\begin{equation}
\gamma_0 = \alpha e^{-\tau m^{\beta}}\label{eq:fit},
\end{equation}
where $\alpha,\beta,\tau$ depend on the parameters of our physical model. 

Finally, we remark that in our calculations, we compute the electrostatic potential and the QD states using high frequency dielectric constants for each material (denoted commonly as $\epsilon_\infty$). This is so because we ignore the time-dependence of a single tunneling event, and thus model tunneling as instantaneous. To be consistent with this viewpoint, or this regime of tunneling, we must take the material response to be of high frequency. Thus the initial and final states of the above matrix element are computed using  $\epsilon_\infty$. We now turn to results of our calculations employing the above model.

\section{Results}\label{sec:results}
\subsection{Potential and wavefunctions}

We begin by plotting the electrostatic potential, and the QD wavefunctions that result from it, to identify the key manifestations of the interplay between confinement (via mass) and the induced polarization. Our choice of parameters in this subsection is made to obtain the  clearest possible visualization of the impact of this interplay. Thus we have selected a relatively low $\epsilon_b = 2.5$, and a barrier potential of 0.8 eV to best capture the qualitative aspects of the self energy and wavefunctions simultaneously within the QD volume and the barrier region. We verify below that the trends in the tunneling rates, based on the physical intuition derived here, continue to hold in more realistic parameter regimes. The distance between the QD and the electrode is chosen such that the effective junction width (the smallest distance between the two surfaces) is about 1 nm. At the microscopic scale, this width may be understood as the distance between the image planes of the two surfaces.

\begin{figure}
\includegraphics[width=1.7in]{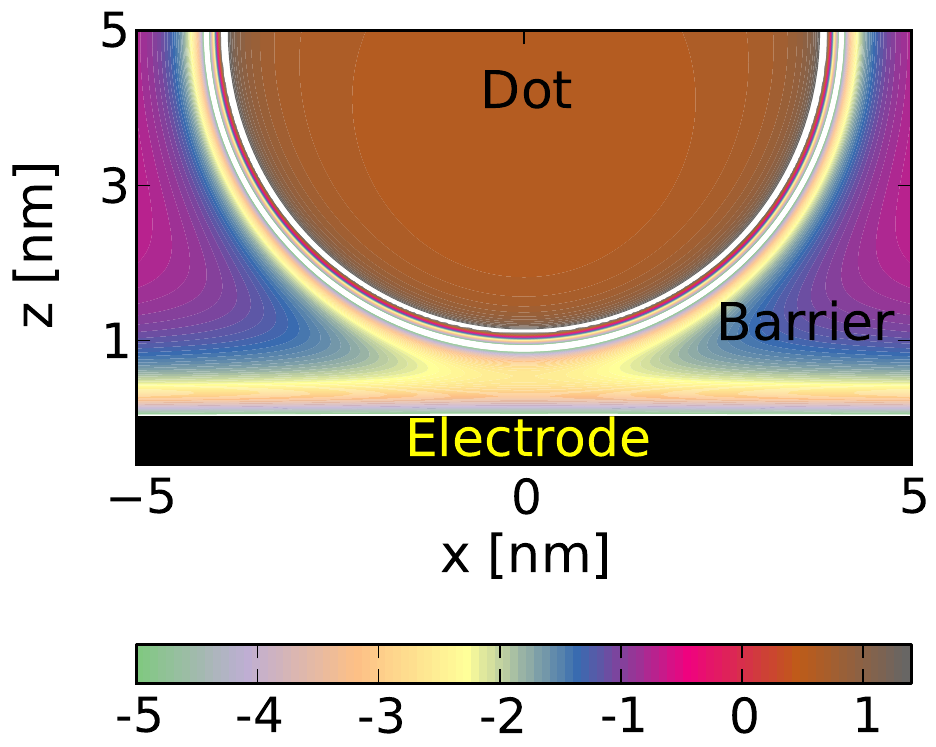}\includegraphics[width=1.7in]{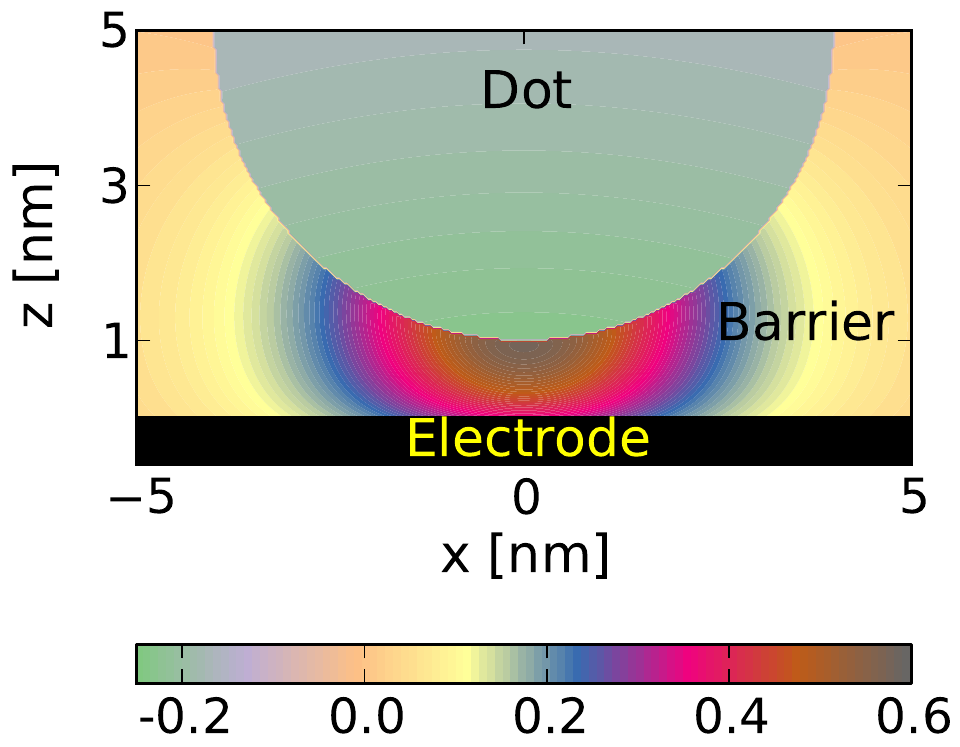}
\vskip 0.5cm
\includegraphics[width=1.7in]{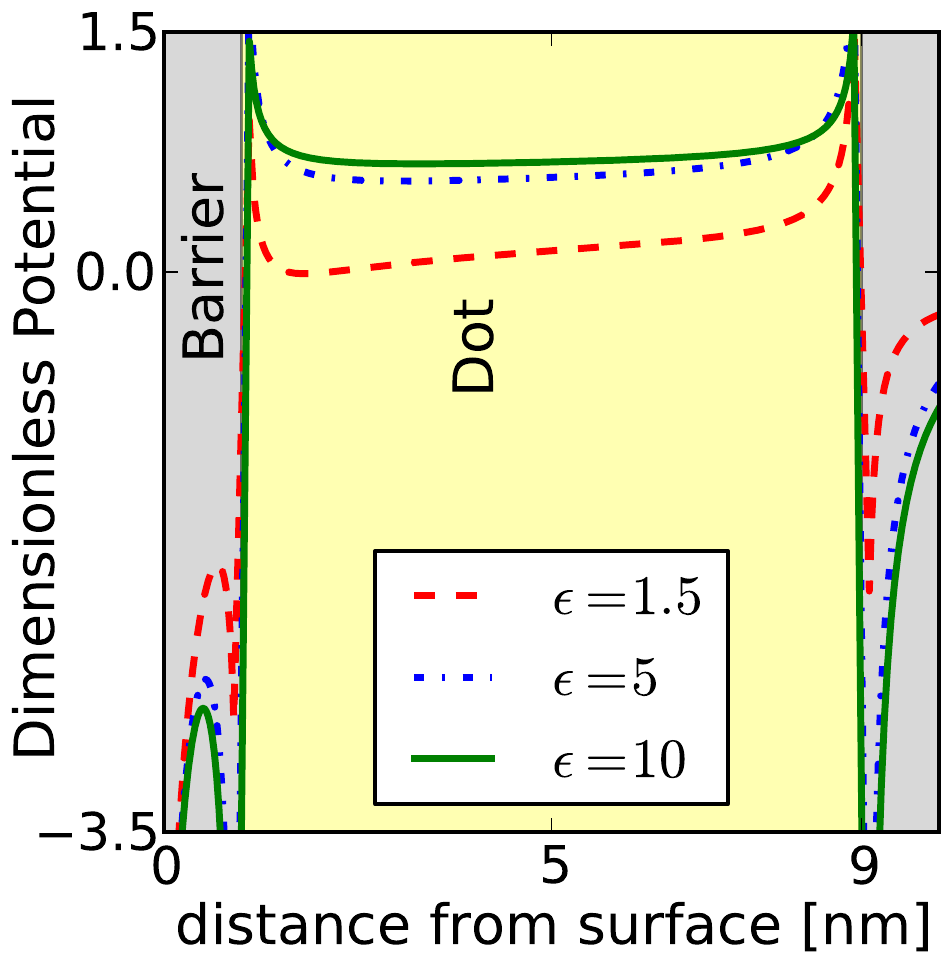}
\caption{\label{fig:Self-energy}Top: Dimensionless self-energy . Upper: (left) contour plot (equatorial plane); (right) correction to multipolar potential induced by the surface. Parameters: metallic electrode  ($f_I=-1$),  $\epsilon=5$, and $\epsilon_b=2.5$. Bottom: Dimensionless self-energy along the $z$ axis for three different $\epsilon=\epsilon_{QD}/\epsilon_b$. }
\end{figure}

Figure \ref{fig:Self-energy} displays the contour plot of dimensionless
self energy in the equatorial plane of the QD, which lies perpendicular
to the electrode surface. We observe that the junction potential is
reduced significantly in a narrow region between the two surfaces. The
reduction is clearer in the lower panel of Fig.~\ref{fig:Self-energy}, where the potential is plotted along the $z$-axis for various dielectric  ratios~$\epsilon$. 

Thus outside the QD, the potential becomes more attractive as $\epsilon$ increases, while inside the QD, the boundary becomes more repulsive. Within the volume of the QD the slope of the potential, resulting from the image charges on the electrode, reduces as a result of dielectric shielding. The narrow attractive potential straddling outside the QD is due to the image potential at the QD boundary, and it becomes asymmetric with greater depth on the side facing the electrode due to the image charges on the electrode. Below, we will see that the asymmetry partially attracts the charge close to the electrode, and shifts the center of mass of the wavefunctions towards the electrode. Therefore, it could significantly affect tunneling, especially for high $\epsilon$ which results in a deep potential well. Since, this attractive potential is beyond the atomic scale variations in the potential at the surface of the QD, we must distinguish it from the surface traps formed at the defects. Therefore, in order to keep the terminology clear, we call this the \emph{image potential trap} (IPT) following Ban\'yai \emph{et al}~\cite{banyai}. 

For quickly estimating the self energy effects, it may be convenient to approximate $\Sigma$ as simply the sum of $\Sigma_0$ and $V_{img}$, the potential of the image charge in~\eqref{eq:Vimg}. The results derived in the previous section allow us to quantify the error in this approximation. In the right hand panel of  Fig. \ref{fig:Self-energy}, we plot this error, which is the contribution by the electrode's correction to self-consistent multipoles corresponding to the total self energy plotted in the left hand panel. It is clear from the plot that the extra contribution is significant only within the junction, and the correction caused by the image effect to the dipolar and higher order field of the QD is mostly quantitative in $\Sigma$ and less than 30\% in most of the space. The size of this correction increases with the QD size as a larger fraction of the charge on its surface becomes close to the electrode and the potential approaches that of a charge between two parallel plates. The contribution decreases as  $\epsilon\rightarrow 1$, and saturates as $(\epsilon-1)/(\epsilon+1)\rightarrow 1$.  This  may be valuable in design considerations involving charged quantum dots. For example, in systems such as epitaxial III-V QDs, the multipole contribution is small, while for inorganic colloidal QDs on planar surfaces and surrounded by vacuum or organic solvents, the contribution could approach its saturation value.

\begin{figure}
\subfloat[]{\includegraphics[width=3.1in]{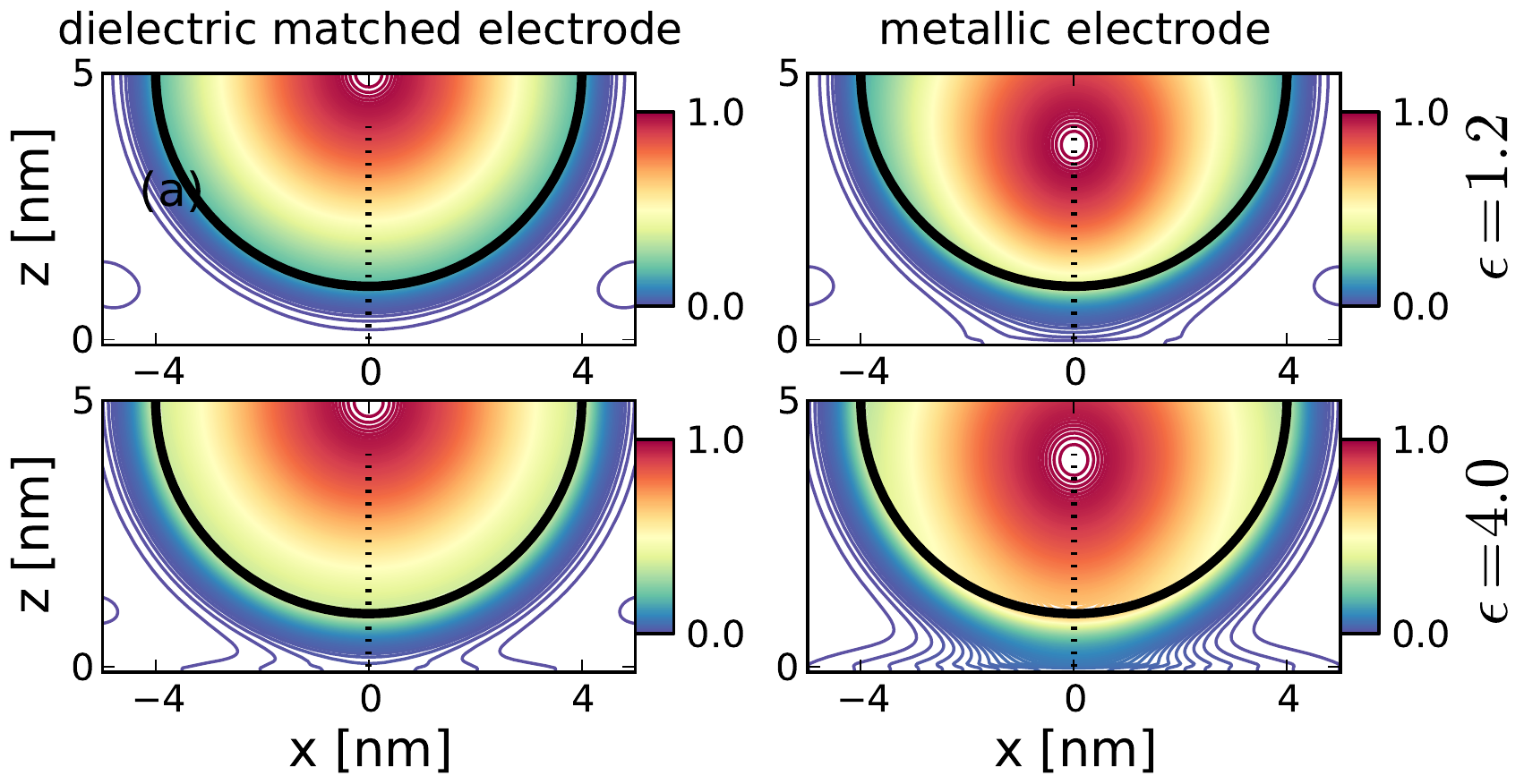}
}

\subfloat[]{\includegraphics[width=2.5in]{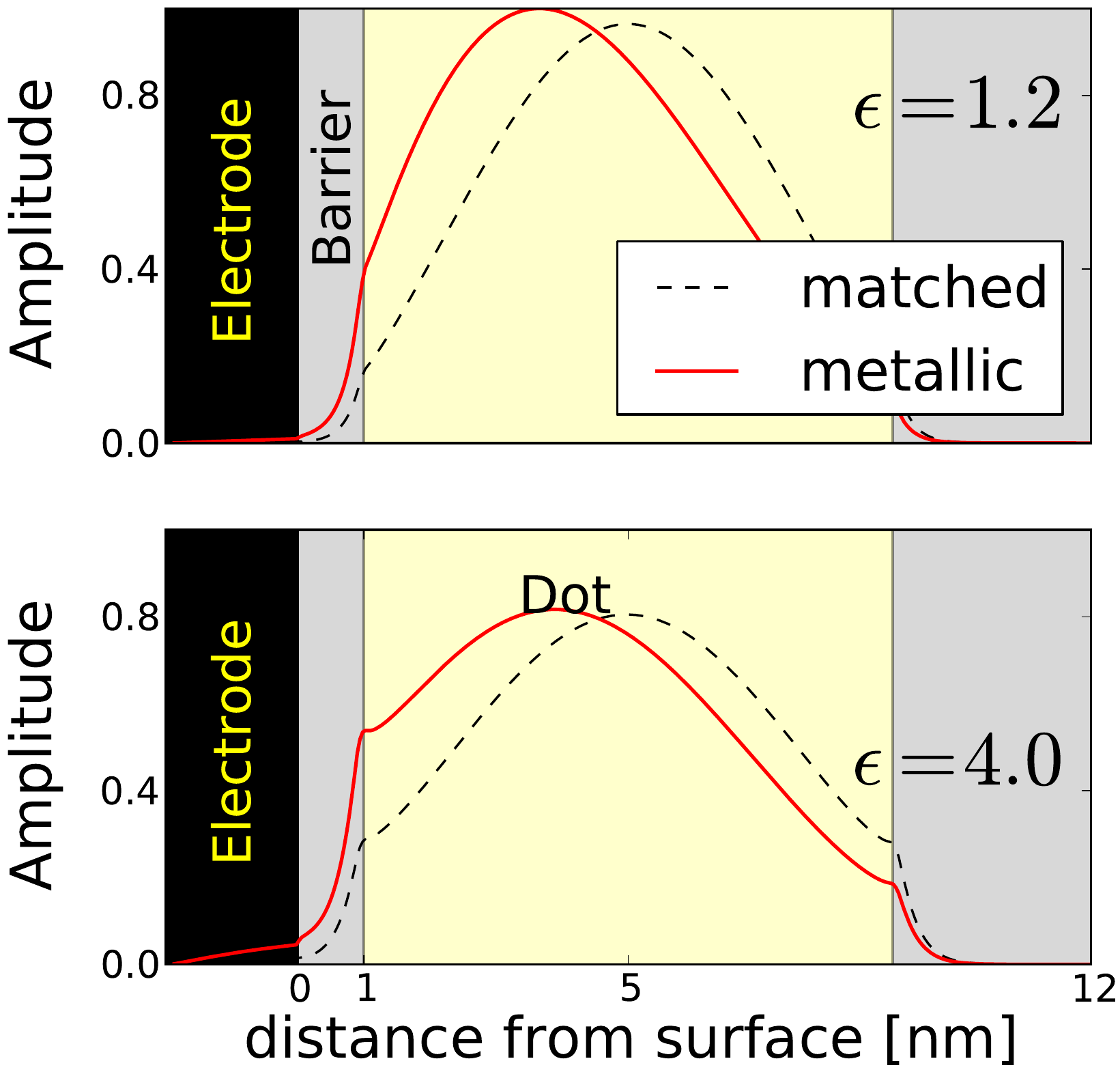}}
\caption{\label{fig:wavefunctions}(a) Wavefunctions in 
the equatorial plane, where thick circle marks the QD boundary, and (b) along the dotted line. In each case, $\epsilon_b=2.5$, $m=0.4$, and $V_b=0.8 $ eV. }
\end{figure}

In Fig. \ref{fig:wavefunctions}(a) we plot the contours of the lowest
energy electron wavefunctions in the equatorial plane for parameter choices corresponding to the two categories in Table~\ref{systemtable} above: low and high values of $\epsilon$. For each $\epsilon$, we plot results at two extreme values of $f_I$: 
matched dielectric at the planar interface boundary with no image effect ($f_I=0$), and a metallic boundary with the full image effect ($f_I=-1$). 
The image effect due to  the metallic boundary leads to increased amplitude in the junction and an overall shift of the center of mass towards the electrode. The center of mass shift is greater at low shielding, which can also be seen in Fig. \ref{fig:wavefunctions}(a) as the center of mass moves back towards the geometrical center of the dot when $\epsilon$ increases from 1.2 to 4.0. At the same time the larger $\epsilon$ yields much higher amplitude outside the QD - a result of stronger barrier lowering and the asymmetry due to the IPT outside the QD. In the lower panel of Fig \ref{fig:wavefunctions}(b), we see that a small peak in the wavefunction forms within the IPT at higher $\epsilon$. Increasing $\epsilon$ further eventually leads to full localization within this region, but at that point our model would lose its physical meaning.

We see here that there is a competition between localization within the QD volume and within the IPT, both of which increase with $\epsilon$. It is the mass of the particle that determines which of these effects wins. The intermediate regime of the two competing effects occurs when the Bohr radius, $4\pi\varepsilon_0\hbar^2e^{-2}(\epsilon_b/m)$, matches the length scale of the IPT. 

The basic effect of the effective mass on localization is discussed and illustrated in the next section (also Fig.~\ref{fig:box-1.0} below), and it can be briefly stated as follows. At low $m$, we expect less localization within both the QD and the IPT. At higher $m$, which is still small enough for significant amplitude outside the QD volume, partial localization in the IPT would increase, which would raise the tunneling rate due to the larger amplitude close to the junction. On the other hand, for sufficiently large $m$, $\psi_D$ would become either highly localized inside the QD volume, or completely within the IPT. Therefore, focusing on the volume states at first, the tunneling rate would be expected to decrease beyond a certain $m$. This suggests that a maximum in tunneling rates must exist, which is confirmed by the calculations below. If we also consider raising $\epsilon$,  the deeper IPT is able to sustain the upward trend in tunneling rates up to  larger $m$, and would thus push the maximum to larger $m$, and saturate the rates as a result of complete trapping on the surface. The latter case is beyond the present model and must include atomic scale contributions to the potential. 

\subsection{Tunneling rates}

We now discuss calculations of tunneling rates for the ground QD states at low barriers (1.0 and 1.5 eV), and for both the lowest and excited states at high barriers (4.5 eV). The low barriers are chosen mainly to illustrate how the physical intuition gained in the previous subsection affects tunneling, while the high barrier cases are more realistic. Note that one can always consider higher energy states to effectively reduce the barrier height, but as discussed below, this leads to complicated profiles of wavefunctions and obscures the trends that exist for the lowest states.

\subsubsection{Low barriers}

\begin{figure}
\includegraphics[width=3.5in]{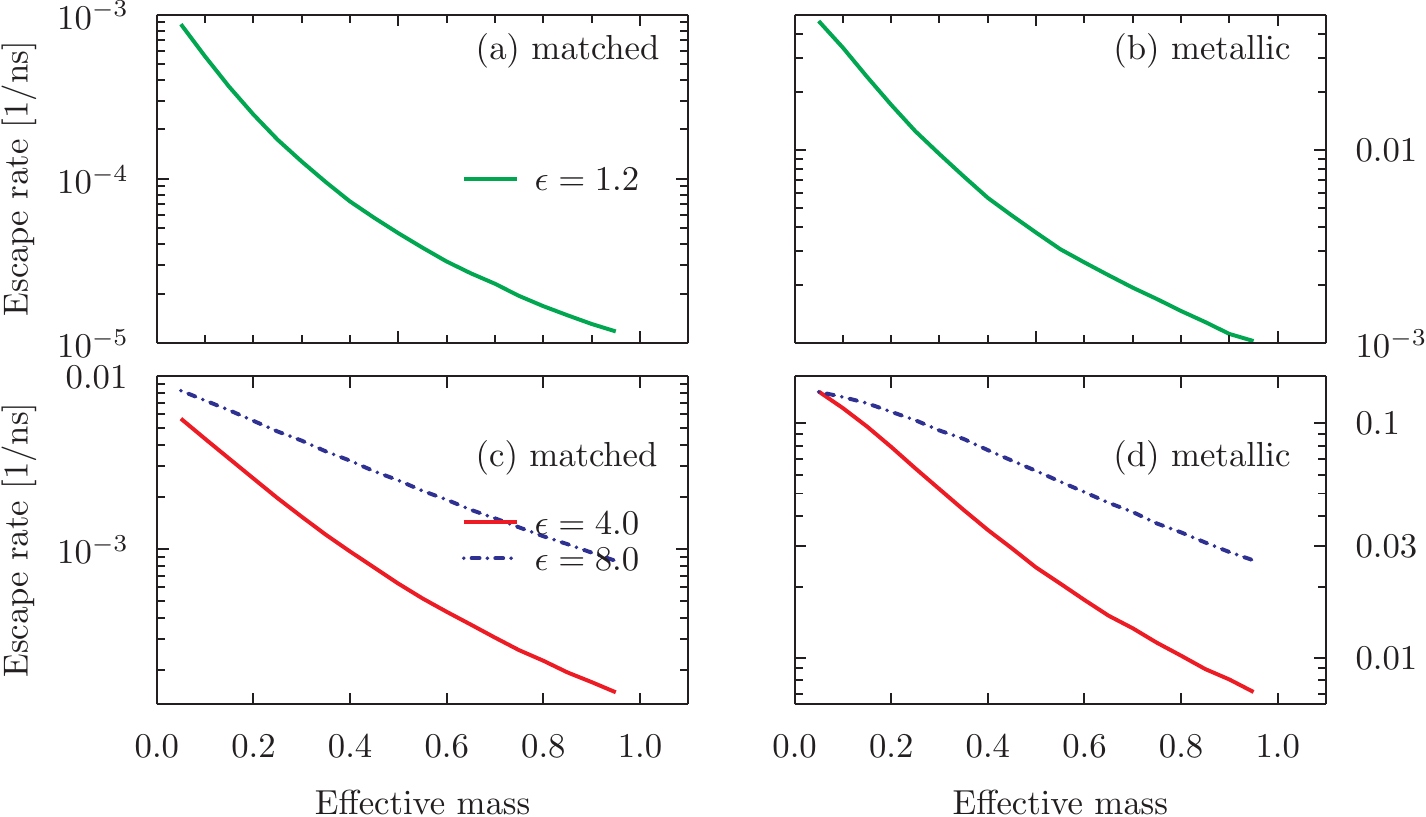}
\caption{\label{fig:box-high}Tunneling rate for metallic and matched dielectric boundary condition at planar interface for barrier height $V_b = 1.5$ eV. Shown are the rates for three different  values of the relative dielectric constants, $\epsilon$ and are normalized to the rate at $m=0.05$ in (a).  The remaining parameters are, barrier dielectric, $\epsilon_b=2.5$, barrier effective mass $m_b=1$, QD radius, $a = 4$ nm, and the distance between the QD center and the electrode, $h/2 = 5$ nm. The rates would scale with the density of states at the electrode, which is taken here to be that of the conduction band of GaAs (parabolic band with effective mass  0.067).}
\end{figure}

\begin{figure}
\includegraphics[width=3.5in,type=pdf,ext=.pdf,read=.pdf]{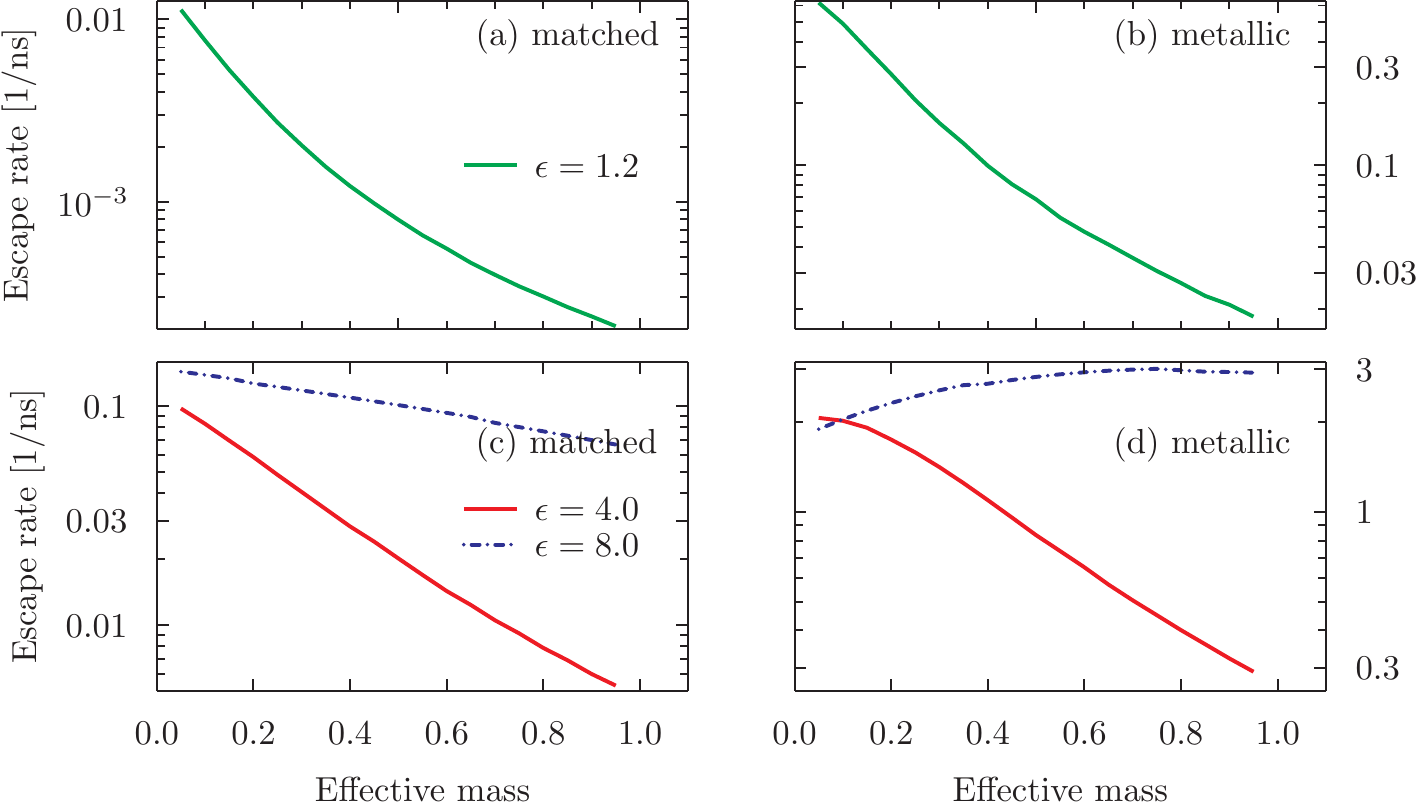}
\caption{\label{fig:box-1.0}Tunneling rate for metallic and matched dielectric boundary condition at planar interface for a barrier height 1.0 eV. The remaining parameters are the same as those in Fig.~\ref{fig:box-high}.}
\end{figure}

\begin{figure}
\includegraphics[width=3.5in]{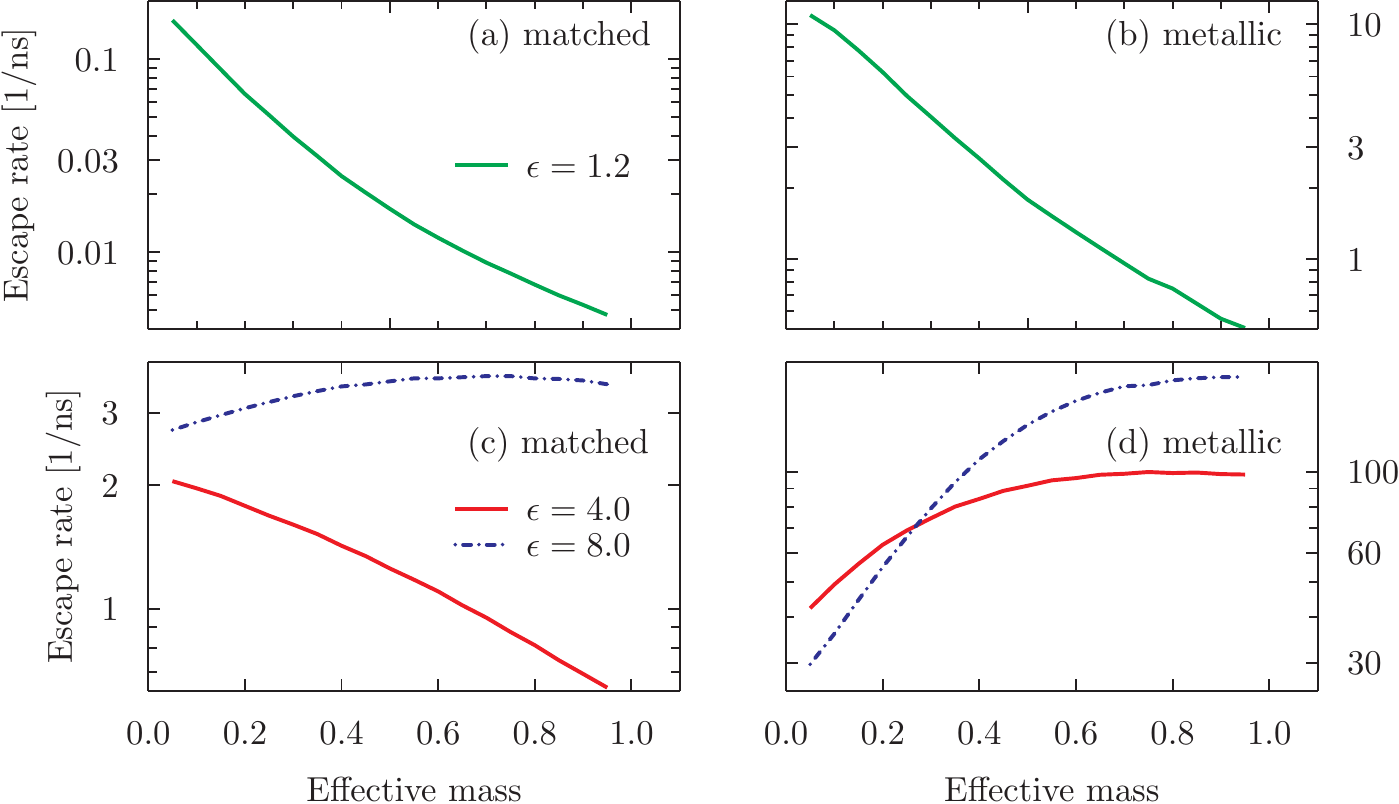}
\caption{\label{fig:box-low}Tunneling rate for metallic and matched dielectric boundary condition at planar interface for a low barrier height of 0.6 eV.  The remaining parameters are the same as those in Fig.~\ref{fig:box-high}. }
\end{figure}

Figures \ref{fig:box-high}-\ref{fig:box-low} show plots of $\gamma_0$ as a function of the effective mass of the particle at low and high values of $\epsilon$, where the figures correspond to barrier heights (in eV) of 1.5, 1.0, and 0.6 respectively. In each figure the value of barrier dielectric constant is $\epsilon_b=2.5$, and the effective mass in the barrier is $m_b=1$. We calculated the rates assuming an isotropic parabolic energy band within the electrode surface having an effective mass of 0.067, which roughly mimics the conduction band of GaAs~\cite{vurgaftman:5815}. The rates would scale in magnitude for different density of states for the electrode, as was discussed previously.

In Fig.~\ref{fig:box-high}, $V_b$ and $\epsilon_b$ are both high enough that  any partial localization in the IPT is suppressed. Recalling our phenomenological expression \eqref{eq:fit} we find that at low ratio, $\epsilon=1.2$,  the rate decays as  $\beta\approx1/2$, and $\tau$ is given approximately by an effective reduction in the junction width due to the image plane of the QD. The deviation in the trend is very slight when the matched electrode is replaced by a metallic one. But the total rate increases by two orders of magnitude due to the significant lowering of the junction potential by the image potential of the electrode. 

However, when $\epsilon$ increases to 4 and beyond, as in panels (c) and (d), the trend also becomes closer to $\beta\approx1$. This is essentially the regime where heavy particle is repelled by the self energy barrier inside the QD much more effectively than the localization by the IPT. 

When we lower the barrier height to 1.0 eV a qualitative change occurs as shown in Fig.~\ref{fig:box-1.0}(d) for $\epsilon=8$: the escape rate is increasing instead of decreasing with an increasing effective mass. In fact, there is a maximum at $m=0.85$ beyond which the rate decreases slowly. This result is the numerical example supporting the discussion at the end of the previous section where we argued that, for sufficiently deep IPTs, the escape rate can increase with mass, or show a maximum.

\begin{figure}[hb]
\includegraphics[width=3.5in]{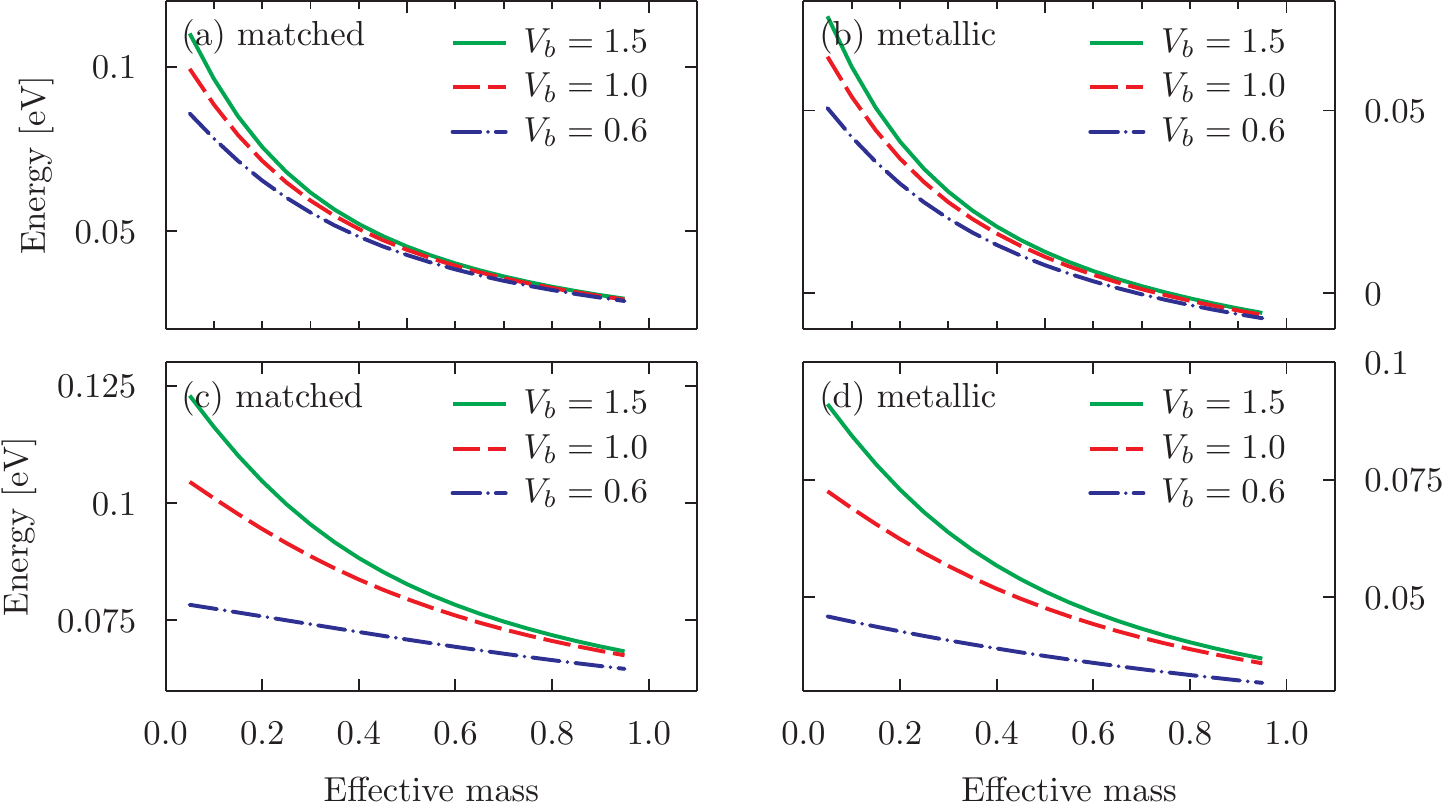}
\caption{\label{fig:energy}Energies of the lowest QD, referenced to the bottom of the well, and for the barrier heights indicated. As in Figures~\ref{fig:box-high}-\ref{fig:box-low} the values of $\epsilon$ are 1.2 in the upper row and 4.0 in the lower row. The remaining parameters are as the same as those in Fig.~\ref{fig:box-high}. Energies at $\epsilon=8.0$ are not included for clarity and follow similar trends as those at 4.0.}
\end{figure}

When the $V_b$ and $\epsilon_b$ are both large enough that the effect of IPT on the wavefunction is negligible, we expect the dependence of $\gamma_0$ on $m$ to follow that of the energy of the lowest state. This can be understood from the binomial expansion of ~\eqref{eq:kappasq}, which gives $\log\gamma_n\approx d\sqrt{2m_bV_b}(1-\Delta E_n/(2V_b)),$ where $d$ is an effective junction width and $\Delta E_n$ is the energy measured from the bottom of the QD potential well. In Fig.~\ref{fig:energy} we plot the energy of the lowest state with respect to the bottom of the potential well as a function of effective mass for three barrier heights. Comparison of the plots in this figure, with those in Figures~\ref{fig:box-high}, \ref{fig:box-1.0}, and \ref{fig:box-low} shows that the trend of monotonically decreasing $\log \gamma_0$ coincides with that of the energy $E_0$ of the QD state. Thus we see that for tunneling, the barrier potential is an important parameter determining the significance of the interplay of electrostatic self energy and effective mass. The barrier dielectric is also significant as it scales the magnitude self energy within the barrier.

\begin{figure}[h]
\includegraphics[width=0.5\columnwidth]{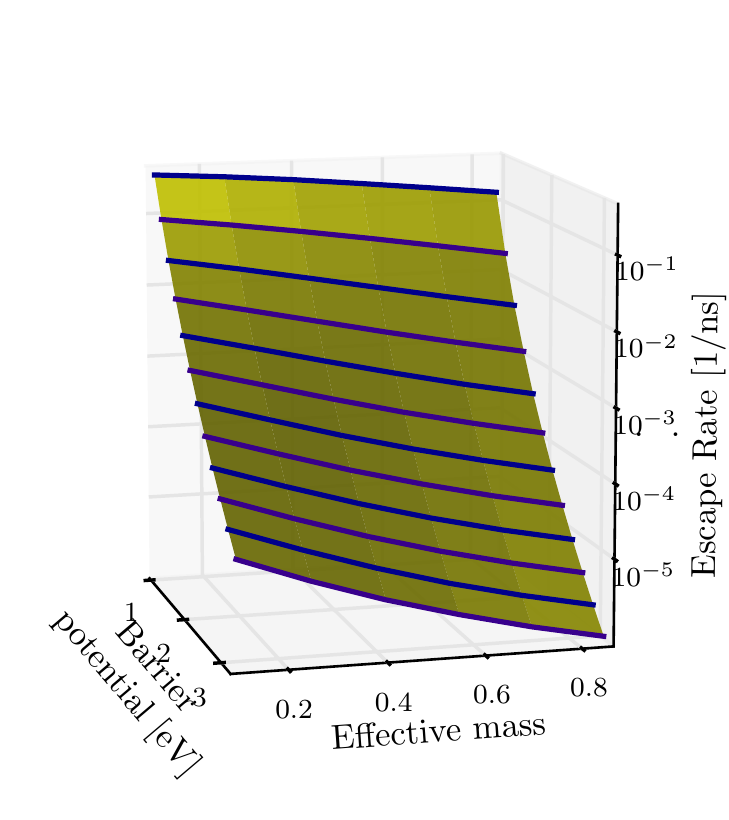}\;\;\includegraphics[width=0.5\columnwidth]{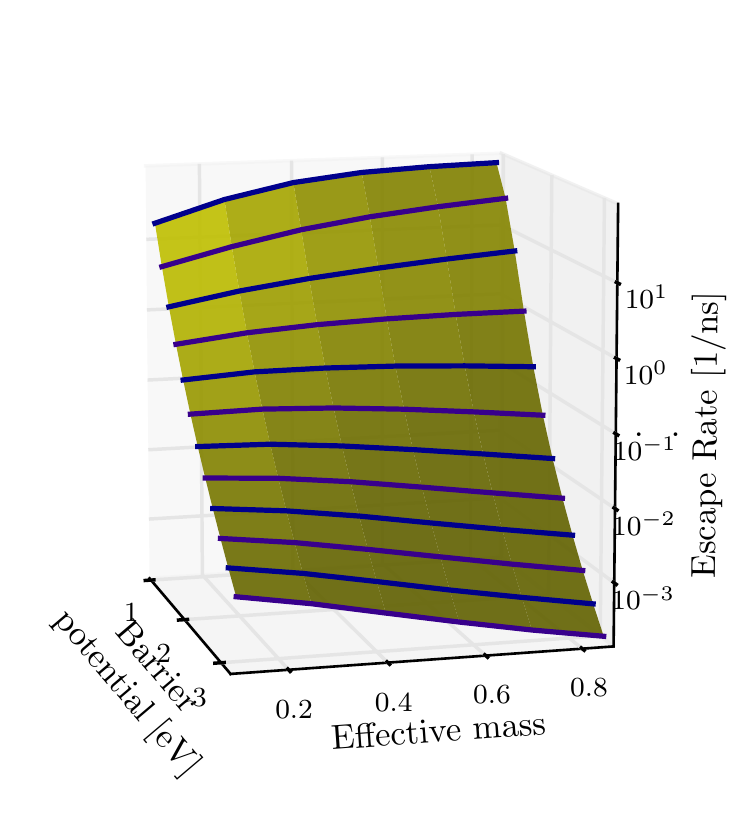}
\caption{\label{fig:3dsheets}Surface plot of escape rates over the two-dimensional space spanned by barrier potential, and effective mass. The lines show the rate as a function of mass as the barrier potential is changed. For the dielectric matched electrode (left), the rate is clearly monotonically decreasing, while for the  metallic electrode (right), the dependence of the rate on mass \emph{changes} from monotonically decreasing to monotonically increasing as the barrier potential is lowered. The remaining parameters are $\epsilon_b=1.6$, $m_b=1$, $\epsilon=8.0$, radius $a = 4$ nm, and $h/2=5$ nm.}
\end{figure}

The most striking result  above is a crossover in the dependence of tunneling rate on the effective mass - switching from decreasing with $m$ to increasing as the barrier height is lowered. In order to visualize the extent of this crossover, we plot in Fig.~\ref{fig:3dsheets} the tunneling rate as a function of mass for a range of barrier potentials  and a dielectric matched as well as a metallic electrode. The existence of crossover can be clearly identified in the figure on the right side.

\subsubsection{High barriers}

It appears from the previous calculations that a potential energy of 1 eV yields a sufficiently high confinement to suppress any interesting behavior such as the crossovers in the tunneling rates. However, in these calculations the barrier dielectric is 2.5. In systems, such as in emerging photovoltaics, the barrier is much closer to a vacuum than to a solid state material, even when QDs in these systems are passivated by organic solvents with low refractive index. In the case of vacuum, the self energy in the barrier increases by almost a factor of 3 compared to its value in the previous calculations, and accounting for both the overall scaling and the increased $\epsilon$ for a fixed QD material. This allows the crossover phenomenon to exist for barrier heights of up to 4-5 eV, which are roughly the work functions of the semiconducting QD materials. 

\begin{figure}[hb]
\includegraphics[width=0.5\columnwidth]{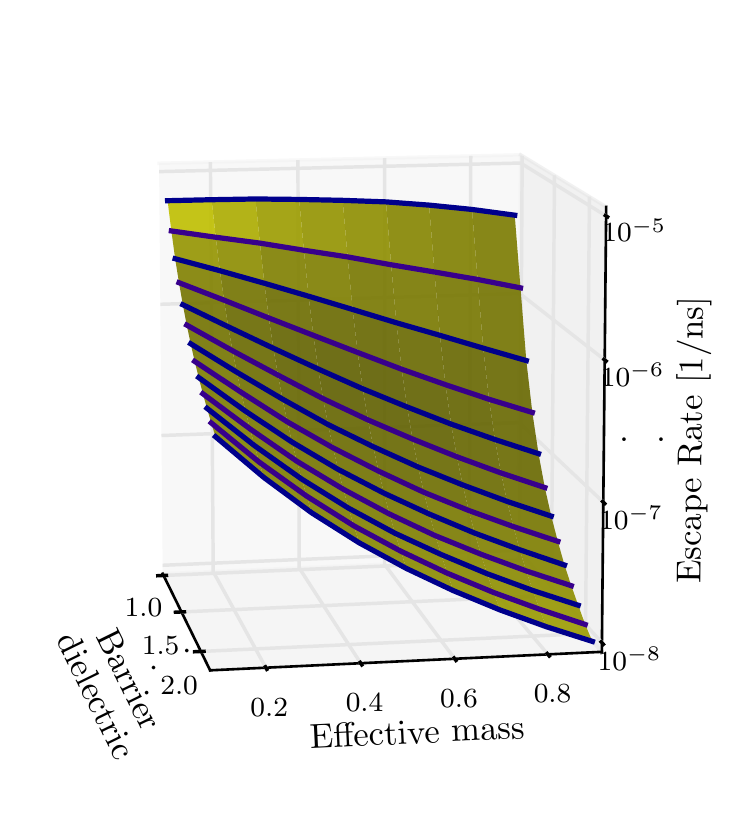}\;\;\includegraphics[width=0.5\columnwidth]{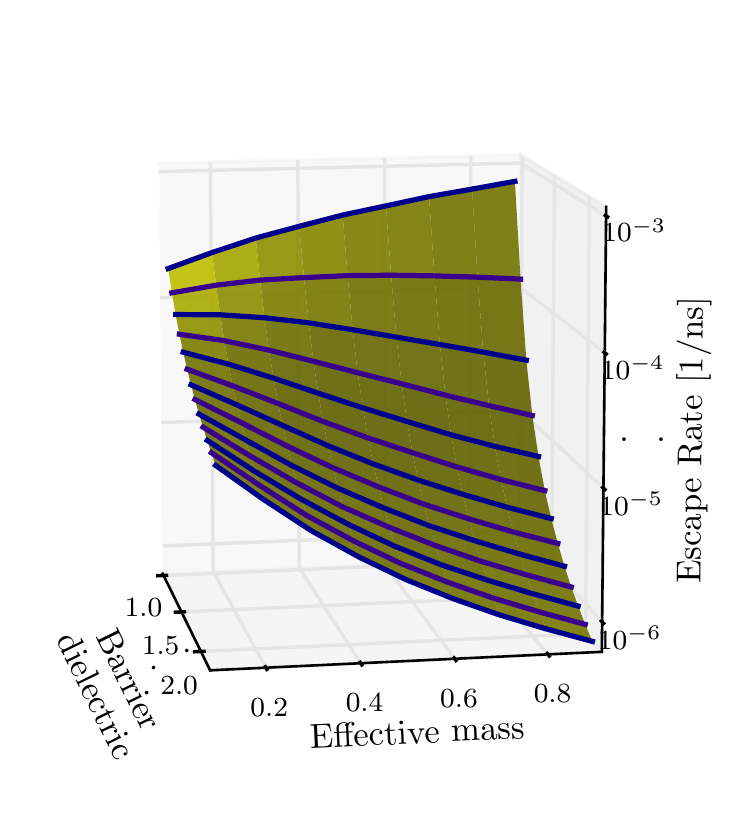}
\caption{\label{fig:crossover-4.5}Surface plot of escape rates over the two-dimensional space spanned by barrier dielectric, and effective mass. The lines show the rate as a function of mass as the barrier potential is changed. For the dielectric matched electrode (left), the rate is clearly monotonically decreasing, while for the  metallic electrode (right), the dependence of the rate on mass \emph{changes} from monotonically decreasing to monotonically increasing as the barrier potential is lowered. The remaining parameters are: barrier height of $V_b=4.5$ eV, $m_b=1$,$\epsilon=8$, radius $a=3$ nm, and $h/2 = 4$ nm.   The effective masses and the barrier heights in these figures may represent systems of II-VI or III-V colloidal quantum dots with the host matrix corresponding to vacuum, p-si, phosphate glass, MgS, or various organic passivating materials (see Fig~\ref{fig:materials}).}
\end{figure}

In Fig.~\ref{fig:crossover-4.5}, we plot the tunneling rates over the plane defined by $(\epsilon_b, m)$ at a fixed $\epsilon = 8.0$. In the absence of image effect by the electrode (left plot), the rate decreases with mass at all $\epsilon_b$, and the decrease becomes more rapid as $\epsilon_b$ increases. However, in the presence of the metallic electrode (right plot), the rate crosses over from decreasing to increasing with mass as $\epsilon_b\rightarrow 1$ from above. In fact, as $\epsilon_b$ decreases, the rate first exhibits a weak maximum, characteristic of the competing effects of localization inside and outside the QD volume, as was discussed previously. The slow rates in these figures are due to the low density of states (DOS) in the electrode model, and as noted earlier, they would be scaled significantly with metallic DOS, and thinner barriers, but without much effect on tunneling rates.

\begin{figure}[h]
\includegraphics[width=3.5in,type=pdf,ext=.pdf,read=.pdf]{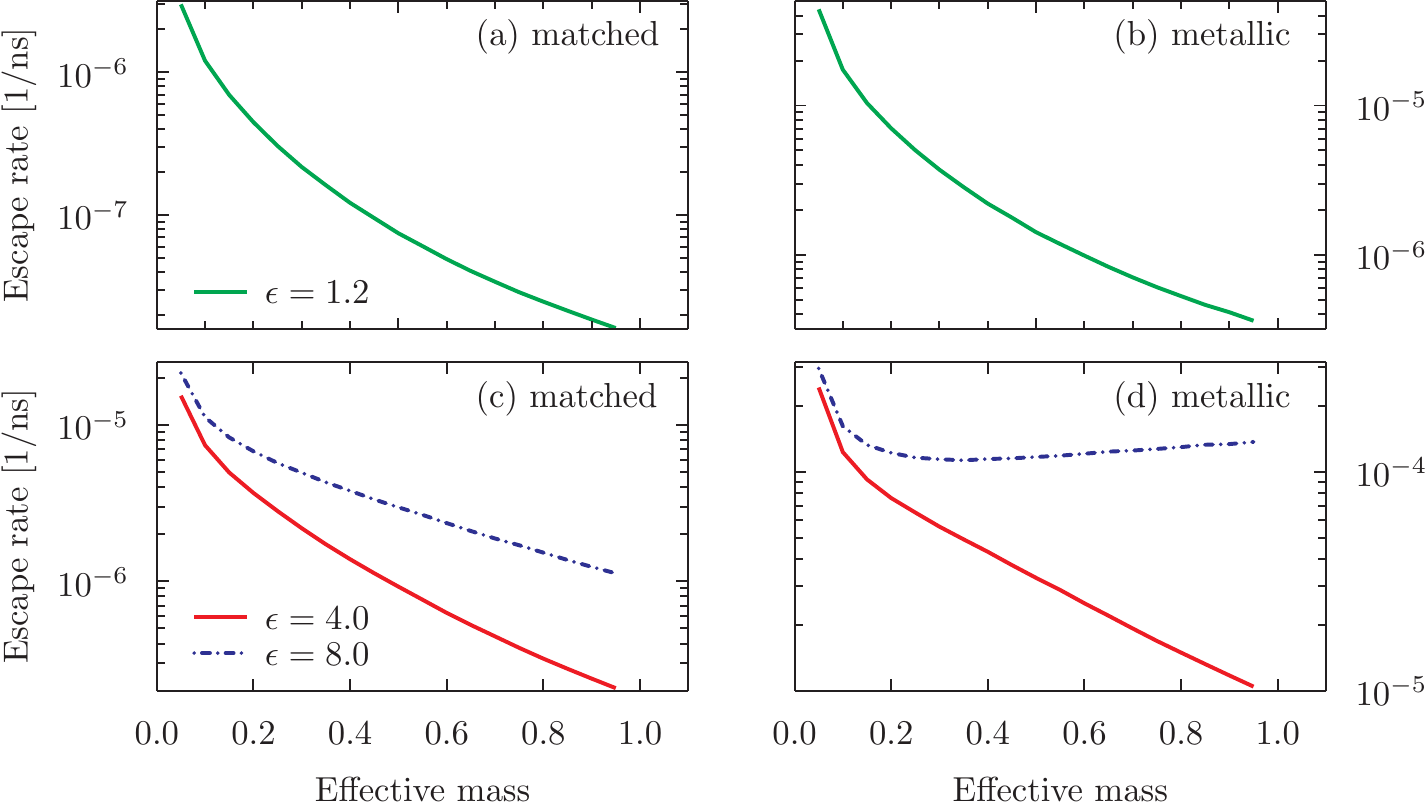}
\caption{\label{fig:box-n1}Tunneling rate for metallic and matched dielectric boundary condition at planar interface for a barrier potential of 4.5 eV, and for the second lowest energy state. Shown are the rates for three different relative dielectric constants, $\epsilon=\epsilon_{QD}/\epsilon_b$, with $\epsilon_b=1.1$. The remaining parameters are the same as in~\ref{fig:crossover-4.5}}
\end{figure}

Finally, we remark that the trends with respect to effective mass also occur for higher energy states. As an example, we show in Fig.~\ref{fig:box-n1} the rates for the second lowest state in the present system. We note for a fixed $\epsilon$ and $m$, there is no substantial difference in the rates with a dielectric matched or a metallic electrode. This is due to the fact that the states involved are deep, and therefore the wavefunction amplitude in the junction is small compared to its values inside the QD. However the effect of the electrode is similar to that in the lowest state, with the characteristic increasing tunneling rate with respect to effective mass inside the QD. The important difference between the two cases is that the rate for higher state shows a minimum rather than a maximum. However the simple interpretation for the trends in the lowest, nodeless, state does not carry over to the excited states directly. This is because the nodes in the excited states act as extra degrees of freedom that determine the wavefunction amplitudes.

At even higher energy, especially close to the top of the potential well of the QD, the wavefunction acquires  a large number of nodes, with a spatial scale reaching below the spatial scale of the image potential. The rates then fluctuate heavily due to the nodal planes moving across the IPT as $\epsilon$ is varied. Thus no obvious trends can be found in this energy range. In addition, microscopic contributions to the potential must be computed in this regime to obtain a robust and reliable trend.


\section{Conclusion}
In summary, we have presented an exact solution to the electrostatic potential of a point charge in a quantum dot lying close to a planar surface. We have found that at low barrier heights (less than 1 eV) there is a significant competition between localization of the particle within the volume of the QD and its localization within the surface trap formed by the image potential at the QD surface. This leads to interesting behavior of the tunneling rate as a function of effective mass, even reversing the trend for larger dielectric contrast between the barrier and the quantum dot. At lower barrier dielectric constant, this trend continues to exist for much higher barriers of up to 4.5 eV.

While the geometry of our model is idealized, the trends may continue to be important in absence of significant geometric perturbations such as protrusions or surface states that may introduce entirely new degrees of freedom relevant to the problem at hand. The basic competition between volume and surface localization will continue to exist in more complex quantum dots as well, and our confirmation of its basic consequences in an idealized model is an important guide to what may occur in those cases. Thus ours is a useful model for an initial attack on a design problem, or an analysis of an experiment in tunneling devices. 

The image effects illustrated here can also be implemented into more complete models at the expense of additional computational effort.  The simple model used here can be generalized to treat two charges, allowing the analysis of the influence of the junction on excitons in a QD. In some cases, the optical dipole moment itself may result from the image force, in which case the multipole fields included here become essential. Finally, the polarization effects discussed here can play an important role in other Coulomb interaction induced phenomena, such as Fermi edge singularity effects in optically excited QDs in close proximity to an electrode~\cite{Kleemans2010}. 

\begin{acknowledgements}
This work is part of the Center for Re-Defining Photovoltaic
Efficiency Through Molecule Scale Control, an Energy
Frontier Research Center funded by the U.S. Department
of Energy (DOE), Office of Science, Office of Basic Energy
Sciences under award no. DE-SC0001085 and the research was carried out in part
at the Center for Functional Nanomaterials, Brookhaven National Laboratory, contract no. DE-AC02-98CH10886. 
KSV also acknowledges partial support by the Natural Sciences and Engineering Research Council of Canada.
\end{acknowledgements}

\bibliographystyle{apsrev4-1}
\bibliography{bibliography}
\newpage
\clearpage

\begin{widetext}
\appendix
\section{Multipole moments}\label{multipole-moments}
\newcommand{\boldr}{\boldsymbol{r}}
\newcommand{\boldh}{\boldsymbol{h}}
\newcommand{\bolds}{\boldsymbol{s}}

In this appendix, we derive the multipole moments, $Q_{lm}$ that appear in the self energy and the two-particle potential in Sec. \ref{sec:theory}. As shown in Fig.~\ref{fig:drawing}, the origin of the cylindrical coordinate system is at the center of the QD. We expand the first two terms of~\eqref{eq:phi} in terms of spherical harmonics, and use the mathematical results on the re-expansion of spherical harmonics with a shifted origin~\cite{caola} to re-write the last term of that equation. The basic relation we use is obtained from Eq. 10 of Caola~\cite{caola}, and using the relationships,
\begin{eqnarray*}
\frac{Y_{lm}(\hat{\boldr})}{r^{l+1}} &=& \sqrt{\frac{(2l+1)(l-m)!}{4\pi(l+m)!}}I_{lm}(\hat{\boldr}),\\
r^lY_{lm}(\hat{\boldr}) &=& \sqrt{\frac{(2l+1)(l-m)!}{4\pi(l+m)!}}R_{lm}(\hat{\boldr}),
\end{eqnarray*}
where the functions $R$ and $I$ are defined in Eq. 1 of Caola~\cite{caola}. With these definitions, we can write, 
\begin{equation}\label{eq:rexp}
\frac{a^{l'+1}}{(r+h)^{l'+1}}Y_{l'm}\left(\frac{\boldr+h\hat{z}}{|\boldr+h\hat{z}|}\right)=\sum_{n=0}^\infty (-1)^{n+m} \gamma_{nm}\gamma_{l'm}C_{n+l'}^{l'} \sqrt{\frac{2l'+1}{2n+1}}\left(\frac{a}{h}\right)^{l'+1}\left(\frac{r}{h}\right)^n Y_{nm}(\hat\boldr),
\end{equation}
where $C_{n+l'}^{l'}=(l'+n)!/(n!l'!)$, and $\gamma_{nm}$ are defined in the main text. The series is well-behaved in the region relevant to the following analysis, which is in the limit $r\rightarrow a$. The last term of~\eqref{eq:phi} now takes the following form,
\begin{eqnarray}
&&\sum_{l'm}\frac{(-1)^{l'+m}Q_{l'm}(\bolds)a^{l'+1}}{(2l'+1)(r+h)^{l'+1}}Y_{l'm}\left(\frac{\boldr+\boldh}{|\boldr+\boldh|}\right)\nonumber\\
&=&\sum_{lm}\sum_{l'}\frac{(-1)^{l'+m}Q_{l'm}(\bolds)}{\sqrt{(2l'+1)(2l+1)}} (-1)^{l+m} \gamma_{lm}\gamma_{l'm}C_{l+l'}^{l'} \left(\frac{a}{h}\right)^{l'+1}\left(\frac{r}{h}\right)^l Y_{lm}(\hat\boldr),
\end{eqnarray}
On the right hand side, we relabeled the index $n$ as $l$ when substituting~\eqref{eq:rexp}. The relabeling makes this term take the form of a coefficient times $Y_{lm}(\hat{\boldr})$, as in the case with the remaining three terms of~\eqref{eq:phi}.

For a charge inside the QD, $s<r<a^{-}$, we take the derivatives of the of potential $\Phi$, defined in \eqref{eq:phi} inside and outside the QD boundary to obtain the displacement field perpendicular to the QD surface. The Maxwell boundary conditions then imply that the components of the displacement field normal to the surface are equal,

\begin{eqnarray*}
\epsilon_b\epsilon\left[\frac{\partial\Phi}{\partial r}\right]_{a^{-}} & =& \frac{e}{a^{2}\varepsilon_{0}} \sum_{lm} Y_{lm}(\hat{\boldsymbol{r}})\left[ 
-\mathfrak{q}_i\frac{\epsilon\left(l+1\right)}{2l+1}\left(\frac{s^{l}}{a^{l}}\right)Y_{lm}^{*}(\hat{\boldsymbol{s}})
+\mathfrak{q}_if_I\frac{\epsilon l}{2l+1}\frac{a^{l}}{\left|\bolds_I\right|^{l}}Y_{lm}^{*}(\hat{\bolds_I})\right.\\
 &  & \left.+\epsilon l\frac{Q_{lm}(\boldsymbol{s})}{2l+1}+f_I\epsilon \sum_{l'}l\frac{Q_{l'm}(\boldsymbol{s})\left(-1\right)^{l'+l}\gamma_{l'm}\gamma_{lm}C_{l'+l}^{l'}}{\sqrt{\left(2l'+1\right)\left(2l+1\right)}}\left(\frac{a}{h}\right)^{l+l'+1}\right]\\
\\
\epsilon_b\left[\frac{\partial\Phi}{\partial r}\right]_{a^{+}} & = & \frac{e}{a^{2}\varepsilon_{0}}\sum_{lm}Y_{lm}(\hat{\boldsymbol{r}})\left[
-\mathfrak{q}_i\frac{l+1}{2l+1}\left(\frac{s^{l}}{a^{l}}\right)Y_{lm}^{*}(\hat{\boldsymbol{s}})
+\mathfrak{q}_if_I\frac{l}{2l+1}\frac{a^{l}}{\left|\bolds_I\right|^{l}}
Y_{lm}^{*}(\hat{\bolds_I})
\right.\\
 &  & \left.-\left(l+1\right)\frac{Q_{lm}(\boldsymbol{s})}{2l+1}
 +f_I\sum_{l'}\frac{lQ_{l'm}(\boldsymbol{s})\left(-1\right)^{l'+l}\gamma_{l'm}\gamma_{lm}C_{l'+l}^{l'}}{\sqrt{\left(2l'+1\right)\left(2l+1\right)}}\left(\frac{a}{h}\right)^{l+l'+1}\right].
\end{eqnarray*}
Anticipating constant values of $\mathfrak{q}(\bolds)$ inside and outside the QD,  we have set here $\mathfrak{q}_i$ as the value of $\mathfrak{q}(\bolds)$ inside the QD. The factor $a^{2}$ in the above equations results from differentiating factors such as $r^{l}/a^{l+1}$,
and then setting $r=a$, which results in $1/a^{2}$. The terms with
$Q_{lm}$ contain an extra $1/a$ in front of the series, and differentiation
is done for $r^{l}/a^{l}$, yielding $1/a$, which combined with the
$1/a$ in front of the series also gives $1/a^{2}$. When we subtract the second equation from the first,  we obtain zero on the left hand side by boundary conditions, and thus the coefficient of each $Y_{lm}$ on the right hand side must vanish. We get, 
\begin{eqnarray*}
0 & = & -\frac{\mathfrak{q}_i\left(\epsilon-1\right)}{2l+1}\left[\left(l+1\right)\left(\frac{s^{l}}{a^{l}}\right)Y_{lm}^{*}(\hat{\boldsymbol{s}})-f_I\frac{la^{l}}{\left|\bolds_I\right|^{l}}Y_{lm}^{*}(\hat{\bolds_I})\right]\\
 &  & +Q_{lm}(\boldsymbol{s})\frac{\left(\epsilon+1\right)l+1}{2l+1}+f_I\left(\epsilon-1\right)\sum_{l'}\frac{lQ_{l'm}(\boldsymbol{s})\left(-1\right)^{l'+l}\gamma_{l'm}\gamma_{lm}C_{l'+l}^{l'}}{\sqrt{\left(2l'+1\right)\left(2l+1\right)}}\left(\frac{a}{h}\right)^{l+l'+1}.
\end{eqnarray*}

When the charge is placed outside outside, we get a similar equation but the one in which only the first term changes, and the resulting equation is
\begin{eqnarray*}
0 & = & \phantom{-}\frac{\mathfrak{q}_o\left(\epsilon-1\right)}{2l+1}\left[l\left(\frac{a}{s}\right)^{l+1}Y_{lm}^{*}(\hat{\boldsymbol{s}})+f_Il\frac{a^l}{\left|\bolds_I\right|^l}Y_{lm}^{*}(\hat{\bolds_I})\right]\\
 &  & +Q_{lm}(\boldsymbol{s})\frac{\left(\epsilon+1\right)l+1}{2l+1}+f_I\left(\epsilon-1\right)\sum_{l'}\frac{lQ_{l'm}(\boldsymbol{s})\left(-1\right)^{l'+l}\gamma_{l'm}\gamma_{lm}C_{l'+l}^{l'}}{\sqrt{\left(2l'+1\right)\left(2l+1\right)}}\left(\frac{a}{h}\right)^{l+l'+1},
\end{eqnarray*}
where $\mathfrak{q}_o$ is the value of $\mathfrak{q}(\bolds)$ for $s>a$.
Writing the two equations together, with Heaviside functions restricting the location of the charge for each, we get, 
\begin{eqnarray*}
 &  & Q_{lm}(\boldsymbol{s})+f_I\sum_{l'}\frac{\left(2l+1\right)}{\left(\epsilon+1\right)l+1}\frac{l\left(\epsilon-1\right)\left(-1\right)^{l'+l}\gamma_{l'm}\gamma_{lm}C_{l'+l}^{l'}}{\sqrt{\left(2l'+1\right)\left(2l+1\right)}}\left(\frac{a}{h}\right)^{1+l+l'}Q_{l'm}(\boldsymbol{s})\\
\\
 & = & 
 \mathfrak{q}(\bolds)\frac{\left(\epsilon-1\right)}{\left(\epsilon+1\right)l+1} 
  \left[\left\{\left(l+1\right)\left(\frac{s^{l}}{a^{l}}\right)\Theta\left(a-s\right)
  -l\left(\frac{a^{l}}{s^{l}}\right)\Theta(s-a)\right\})
  Y_{l}^{m*}(\hat{\boldsymbol{s}})
  -f_Il\frac{a^{l}}{\left|\boldsymbol{s}_I\right|^{l}}Y_{lm}^{*}(\hat{\bolds_I})\right].
\end{eqnarray*}
The first term in the right hand side of this equation defines the $Q_{lp}^{(0)}(\boldsymbol{s})$ in the main text, which also generate the third term, and the coefficients of $Q_{lm}$ on the left hand side define the matrix elements $A_{ll'}^m$, both of which are used in Sec.~\ref{sec:theory} to write the expressions for $\Sigma$ and $\Phi$.
 
\section{Self energy regularization}\label{regularize}
 
 When we put $\boldsymbol{r=}\boldsymbol{s}$ in $\Phi(\boldsymbol{r};\boldsymbol{s})$,
the resulting formula has two terms arising from the zeroth order contribution $Q^{(0)}$, where one term is for $s<a$ and the other for $s>a$. They can both be written in a single expression as,
\begin{eqnarray*}
 &  & \sum_{l=0}^{\infty}\frac{\left(l+\beta\right)}{\left[\left(\epsilon+1\right)l+1\right]}\rho^{2l},
\end{eqnarray*}
where $\beta=1$ corresponds to $s<a$ and $\beta=0$ to $s>a$, (with prefactors not relevant here). In each of the two
cases, the series at large $l$ becomes a harmonic series and therefore
divergent. To isolate the divergent terms, we re-write the above series
as 
\begin{eqnarray}
 & = & \beta+\sum_{l=1}^{\infty}\left[\frac{\left(l+\beta\right)}{\left[\left(\epsilon+1\right)l+1\right]}-\frac{1}{\left(\epsilon+1\right)}-\frac{\alpha}{\left(\epsilon+1\right)^{2}(l+1)}\right]\rho^{2l}
  +\frac{1}{\epsilon+1}\sum_{l=1}^{\infty}\rho^{2l}+\frac{\alpha}{\left(\epsilon+1\right)^{2}}\sum_{l=1}^{\infty}\frac{\rho^{2l}}{l+1},\label{eq:sumterm}
\end{eqnarray}
where $\alpha$ is to be determined. The last two terms of the above
equation can be summed to obtain,
\begin{eqnarray*}
\frac{1}{\epsilon+1}\sum_{l=1}^{\infty}\rho^{2l} & = & \frac{1}{\left(\epsilon+1\right)}\frac{\rho^{2}}{1-\rho^{2}}\\
\frac{\alpha}{\left(\epsilon+1\right)^{2}}\sum_{l=1}^{\infty}\frac{\rho^{2l}}{l+1} & = & -\frac{\alpha}{\left(\epsilon+1\right)^{2}}\left[\frac{1}{\rho^{2}}\ln\left(1-\rho^{2}\right)+1\right].
\end{eqnarray*}
The coefficient of the $\rho^{2l}$ term in the first sum on the right
hand side of \ref{eq:sumterm} is,

\begin{eqnarray*}
 &  & \frac{1}{\left(\epsilon+1\right)^{2}}\left[\frac{\left(\epsilon+1\right)\left(l+1\right)\left[\beta\left(\epsilon+1\right)-1\right]-\alpha\left(\epsilon+1\right)l-\alpha}{\left[\left(\epsilon+1\right)l+1\right]\left(l+1\right)}\right],
\end{eqnarray*}
which behaves as $1/l$ at large $l$, unless we set the coefficient
of $l$ in the numerator equal to zero. This condition determines
the value of $\alpha$ and yields $\alpha  =  \beta\left(\epsilon+1\right)-1$. The numerator is then $\epsilon\left[\beta\left(\epsilon+1\right)-1\right]$.
Thus 
\begin{eqnarray*}
\sum_{l=0}^{\infty}\frac{\left(l+\beta\right)}{\left[\left(\epsilon+1\right)l+1\right]}\rho^{2l} & = & \beta+\frac{\epsilon\left[\beta\left(\epsilon+1\right)-1\right]}{\left(\epsilon+1\right)^{2}}\sum_{l=1}^{\infty}\frac{\rho^{2l}}{\left[\left(\epsilon+1\right)l+1\right]\left(l+1\right)}\\
 &  & +\frac{1}{\left(\epsilon+1\right)}\frac{\rho^{2}}{1-\rho^{2}}-\frac{\beta\left(\epsilon+1\right)-1}{\left(\epsilon+1\right)^{2}}\left[\rho^{-2}\ln\left(1-\rho^{2}\right)+1\right].
\end{eqnarray*}
Substitution of $\beta=0$ or $\beta=1$ yields the self energy series shown in the main text.

\end{widetext}

\end{document}